\documentstyle[12pt]{article}
  \let\bsk=\bigskip \let\qd=\quad
\let\a=\alpha \let\be=\beta \let\g=\gamma \let\de=\delta
\let\ep=\varepsilon  \let\et=\eta \let\Th=\theta
  \let\la=\lambda \let\m=\mu
\let\n=\nu  \let\r=\rho \let\si=\sigma
\let\om=\omega 
\let\ph=\varphi  \let\PH=\Phi 
\let\Om=\Omega  
 \let\De=\Delta

\def\lb{\left(} \def\rb{\right)}
\def\0#1#2{\frac{#1}{#2}}
\def\s0#1#2{\mbox{\small{$\frac{#1}{#2}$}}}
\def\5{\bar }  \def\6{\partial } \def\7{\hat } \def\4{\tilde }
\let\LRA=\Leftrightarrow \let\then=\Rightarrow

\let\nn=\nonumber \let\lra=\leftrightarrow
\def\simple{\, \stackrel{\mbox{\Gl{xx1}}}{=}\, }
\def\bea{\begin{eqnarray}} \def\eea{\end{eqnarray}}
\def\beann{\begin{eqnarray*}} \def\eeann{\end{eqnarray*}}
\def\beq{\begin{equation}} \def\eeq{\end{equation}}
\def\ba{\begin{array}} \def\ea{\end{array}}
 \def\cB{{\cal B}} \def\cG{{\cal G}}
\def\cA{{\cal A}}  \def\cL{{\cal L}}
  \def\cC{{\cal C}}
\def\cF{{\cal F}} \def\cT{{\cal T}} 
 \def\cP{{\cal P}} \def\cS{{\cal S}}
 \def\cE{\7\Theta}

\def\oor{{\5r}} 
 
 \def\R#1#2#3{{R_{#1#2}}^{#3}}
\def\F#1#2#3{{F_{#1#2}}^{#3}} \def\G#1#2#3{{\Gamma_{#1#2}}^{#3}}
 \def\f#1#2#3{{f_{#1#2}}^{#3}}

\def\e#1#2{{e_{#1}}^{#2}} \def\E#1#2{{E_{#1}}^{#2}}
 \def\A#1#2{{A_{#1}}^{#2}}
\def\As#1#2{A_{#1}^{*#2}} \def\hs#1#2{e_{#1}^{*#2}}
\def\o#1#2{{\om_{#1}}^{#2}}

\newcommand{\mysection}[1]{\section{#1}\setcounter{equation}{0}
\setcounter{theorem}{0}\setcounter{lemma}{0}\setcounter{corollary}{0}}
\def\Gl#1{(\ref{#1})}
\def\ben{\begin{enumerate}}
\def\een{\end{enumerate}}
\def\act{\cS} 
\def\ainv{{\cal A}_{2,0}^{inv}}
\def\qed{\hbox{${\vcenter{\vbox{
   \hrule height 0.4pt\hbox{\vrule width 0.4pt height 6pt
   \kern5pt\vrule width 0.4pt}\hrule height 0.4pt}}}$}}
\newtheorem{theorem}{Theorem}
\newtheorem{lemma}{Lemma}

\newtheorem{corollary}{Corollary}
\newcommand{\proof}[1]{{\bf Proof.} #1~$\qed$}

\def\oor{{\5r}} 
 \def\gh#1{gh(#1)} \def\agh#1{antigh(#1)}
\newcommand{\LAB}{{(K)}}
\newcommand{\La}{\alpha}
\newcommand{\Lb}{\beta}
\newcommand{\Ly}{m}

\begin{document}
\thispagestyle{empty}
\hspace*{\fill} KUL--TF--95/16\\
\hspace*{\fill} ULB--TH--95/07\\
\hspace*{\fill} hep-th/9505173\\
\vfill

{\renewcommand{\thefootnote}{\fnsymbol{footnote}}
\addtolength{\footnotesep}{1ex}
\begin{center}
{\LARGE Local BRST cohomology in\\[1ex]
Einstein--Yang--Mills theory}
\end{center}
\vfill

\begin{center}
{\Large
Glenn Barnich$^{1,}$\footnotemark,
Friedemann Brandt$^{2,}$\footnotemark
\\ and
Marc Henneaux$^{1,}$\footnotemark}
\end{center}
\vfill
\ben
\item[$^1$] Facult\'e des Sciences, Universit\'e Libre de Bruxelles,
Campus Plaine C.P. 231, B--1050 Bruxelles, Belgium
\item[$^2$] Instituut voor Theoretische Fysica, Katholieke
Universiteit Leuven,
Celestijnenlaan 200 D, B--3001 Leuven, Belgium
\een
\vfill
\addtocounter{footnote}{-3}
\addtocounter{footnote}{1}
\footnotetext{Aspirant au Fonds National de la
Recherche Scientifique (Belgium).\\ E-mail address: gbarnich@ulb.ac.be}
\addtocounter{footnote}{1}
\footnotetext{Supported by the research council (DOC) of the K.U. Leuven.\\
E-mail address: Friedemann.Brandt@fys.kuleuven.ac.be}
\addtocounter{footnote}{1}
\footnotetext{Also at Centro de Estudios
Cient\'\i ficos de Santiago, Chile.\\ E-mail address: henneaux@ulb.ac.be}
}
\newpage
\thispagestyle{empty}

\begin{abstract}
We analyse in detail the local BRST cohomology in Einstein-Yang-Mills theory
using the antifield formalism. We do not restrict the Lagrangian to be the sum
of the standard Hilbert and Yang-Mills Lagrangians, but allow for more general
diffeomorphism and gauge invariant actions. The analysis is carried out in all
spacetime dimensions larger than 2 and for all ghost numbers. This covers the
classification of all candidate anomalies, of all consistent deformations of
the action, as well as the computation of the (equivariant) characteristic
cohomology, i.e. the cohomology of the spacetime exterior derivative in the
space of (gauge invariant) local differential forms modulo forms that vanish
on-shell. We show in particular that for a semi-simple Yang-Mills gauge group
the antifield dependence can be entirely removed both from the consistent
deformations of the Lagrangian and from the candidate anomalies. Thus, the
allowed deformations of the action necessarily preserve the gauge structure,
while the only candidate anomalies are those provided by previous works not
dealing with antifields, and by ``topological" candidate anomalies which are
present only in special spacetime dimensions (6,9,10,13,...). This result no
longer holds in presence of abelian factors where new candidate anomalies and
deformations of the action can be constructed out of the conserved Noether
currents (if any). The Noether currents themselves are shown to be
covariantizable, i.e. they can be chosen to be invariant under local Lorentz
and Yang-Mills transformations and covariant under diffeomorphisms, with a
few exceptions discussed as well.
\end{abstract}
\newpage
\setcounter{page}{1}
\setcounter{footnote}{0}
\mysection{Introduction}\label{intro}

\subsection{Motivation}

This paper deals with the following physical questions that
arise in $n$-di\-men\-sional
Einstein--Yang--Mills theory (as they actually arise in
any local field theory).
\ben
\item[(i)] Characteristic cohomology \cite{Bryant}:\\
Do there exist non-trivial local spacetime $p$-forms constructed
out of the fields and their derivatives that are conserved, i.e.
closed modulo the equations of motion? A conserved $p$-form is
said to be trivial if it is weakly exact, i.e. exact on-shell.
In particular, what is the structure of the conserved local
currents ($p=n-1$)? Are there non-trivial conserved $(n-2)$-forms
(``charge without charge" in Wheeler's terminology \cite{Wheeler,Unruh})?
Of particular interest are these questions in the restricted space
of gauge invariant local forms (``equivariant characteristic
cohomology").
\item[(ii)] Consistent deformations of the action \cite{hen}:\\
Can one add non-trivial ``interaction terms" to the action in a
way that maintains the number (but not necessarily the form
nor the group structure) of the gauge symmetries? Such
deformations are called ``consistent" and are regarded as trivial
if they arise from a mere local redefinition of the fields.
The consistent deformations determine the possible quantum
corrections to the action (in the absence of anomalies).
Well-known solutions are given by invariant functions
of the curvatures, the matter fields and their covariant
derivatives. They do not change the gauge structure.
It is of interest to determine whether there are other deformations.
\item[(iii)] Candidate anomalies:\\
What are the possible anomalies in the quantum theory, i.e.
what is the structure of the possible breakings of the gauge
symmetries by quantum effects?
\een
These important questions have already received quite a lot of
attention in the past, under restrictive assumptions. For
instance, in \cite{Unruh} Unruh has shown the absence of
``charges without charges" in Einstein gravity that would be
at most of degree 5 in derivatives (of the metric) and
covariant, i.e. constructable out of the undifferentiated
metric, the Riemann tensor and its covariant derivatives.
Similarly, the problem of consistent deformations of the Einstein theory
has been treated by Wald in his analysis of possible spin-2
theories \cite{Wald}, but always under the assumption that the algebra
of the deformed gauge symmetries still closes off-shell (under commutation).
Finally, the classification of the anomalies has been discussed
at length in \cite{gravlit,grav}, where complete results
on the candidate anomalies not containing the sources for
the BRST variations (antifields) are given. The purpose of
this paper is to re-address the issues (i)--(iii) above
without making any a priori simplifying assumptions on the
solutions or the spacetime dimension (except that we assume the
latter to be larger than 2).

\subsection{Method}

Our approach  is entirely based on the reformulation of (i)--(iii)
in terms of the Wess--Zumino consistency condition
when rewritten in
BRST language, extended to arbitrary ghost number and
form degree and taking into account the equations of motion.
The Wess--Zumino consistency condition then reads
\beq s a^{g,p}+d a^{g+1,p-1}=0\label{i1}\eeq
where $s$ is the Becchi--Rouet--Stora--Tyutin (BRST) operator
\cite{brs}--\cite{bv}
given explicitly below and where $d$ is the
spacetime exterior derivative. The local forms $a^{i,j}$
have ghost number $i$ and form degree $j$ where $i$ takes integer
values (including negative ones) and
$0\leq j \leq n$. They involve both the fields, the ghosts,
the antifields and their derivatives. Trivial
solutions of \Gl{i1} are given by $ a^{g,p}= sb^{g-1,p}+db^{g,p-1}$ and
satisfy indeed \Gl{i1} due to
\beq s^2=sd+ds=d^2=0.\label{nili}\eeq
Equation \Gl{i1} defines
the local BRST cohomological group $H^{g,p}(s|d)$ as
the set of equivalence classes of solutions $a^{g,p}$ of
\Gl{i1} modulo trivial ones.

In \cite{wz} Wess and Zumino showed that candidate
anomalies must fulfill \Gl{i1} for $g=1$ and $p=n$. It is easy
to verify that consistent deformations of the action are
solutions of \Gl{i1} with $g=0$ and $p=n$ \cite{hen}.
Perhaps less known is the fact that the characteristic
cohomology is also described by the
Wess--Zumino consistency condition, but this time at negative
ghost number. This was proved in \cite{bbh1}. Consequently,
the issues (i)--(iii) are indeed equivalent to solving
the Wess--Zumino consistency condition for values of the
ghost number $\leq 1$, which provides a unified approach to
these questions.

\subsection{Main results}\label{1.3}

We indicate in this paper how to compute the BRST cohomology
groups $H^{g,p}(s|d)$ for Einstein--Yang--Mills theory, and
give the form of the most general cocycle (for all values of
the ghost number and the form degree).
Our main results, which were partly announced
in \cite{rapid}, can be summarized as follows.
\bsk

\noindent (i) Characteristic cohomology:

The conserved $p$-forms ($p\leq n-1$) are of two types.

\noindent (i.a) Topological conserved $p$-forms.

These are closed (but non-exact) $p$-forms $a^p$ which fulfill
\[ da^p=0,\qd a^p\neq db^{p-1}\]
identically, without invoking the equations of motions.
Their existence (on a spacetime manifold with {\bf R}$^n$ topology)
follows from the non-triviality of the vielbein manifold.
They generalize the conservation laws found by Finkelstein and Misner
\cite{Fink}. The topological conserved $p$-forms may be chosen to
involve only the vielbeins and their exterior derivatives
and can be described in terms of the Lie algebra cohomology
of $so(n)$.

\noindent (i.b) Dynamically conserved $p$-forms.

Conserved $p$-forms of the second type are only weakly closed,
\[
da^p\approx 0,\qd a^p\not\approx db^{p-1}
\]
where $\approx$ denotes weak equality, i.e. equality
up to terms that vanish on-shell, and
it is understood that all topological conserved $p$-forms
have been removed from $a^p$.
Dynamically conserved $p$-forms
can be expressed entirely in terms of objects that transform
tensorially under the gauge symmetries. For $p<n-2$, there are
no non-trivial dynamically conserved $p$-forms.
For $p=n-2$, there is one conserved $(n-2)$-form for each
``free" abelian Yang--Mills gauge field (i.e. for each
abelian factor of the gauge group under which all
the matter fields are uncharged),
and these forms constitute a complete set of
inequivalent dynamically conserved $(n-2)$-forms.
They correspond to the ``charge without charge" mechanism
of Wheeler. Finally, while the characteristic
cohomology for $p\leq n-2$ is relatively insensitive
to the form of the Lagrangian, this is not the case for
$p=n-1$. According to Noether's first theorem \cite{Noether}
there may exist dynamically conserved currents if the action
has global symmetries, e.g. baryon number symmetry. This
depends in particular on the matter part of the
action (the absence of non-trivial conserved currents for
pure $n=4$ Einstein gravity has been demonstrated in \cite{anto}).
Since we do not restrict our investigation to a particular
action, we do not investigate this question in detail, but
rather prove that all Noether currents may be chosen
to be invariant under local Lorentz and Yang--Mills gauge
transformations and covariant under diffeomorphisms,
with a few exceptions discussed in the text.

For the equivariant characteristic cohomology we prove
that its cohomology classes
are represented by the covariantized non-trivial
Noether currents (written as
$(n-1)$-forms), by $(n-2)$-forms related with ``free"
abelian gauge fields and by invariant polynomials in the
curvature 2-forms. In particular, the latter are
shown to be non-exact (even on-shell) in the
space of invariant $p$-forms for $p<n-1$. This holds also in degree
$(n-1)$ (relevant only in odd dimensions) if
``free" abelian gauge fields are absent, and in degree $n$ (in
even dimensions) it holds in absence of
non-trivial Noether currents containing Chern--Simons terms.
\bsk

\noindent (ii) Consistent deformations of the action:

Ghost number zero solutions of the
Wess--Zumino consistency condition independent of the
antifields define deformations of the action that do not
modify the gauge symmetry. Solutions linear in the
antifields change the gauge symmetry, but in such a
way that the algebra of the new gauge transformations still closes
off-shell. Solutions quadratic (or of higher order) in the
antifields modify more radically the gauge symmetry, since the
algebra of the new gauge transformations ceases to close
off-shell.

Our main result here is that the most general solution
$a^{0,n}$ of \Gl{i1} can always be redefined such that it
does not involve the antifields, provided the Yang--Mills gauge group
is semi-simple. Hence, in that case the form of the gauge transformations
can be kept intact under any deformation of the action.
This justifies the assumptions made by Wald, and implies that
the only consistent deformations are given by scalar densities
constructed out of the curvatures,
the matter fields and their covariant derivatives and the
undifferentiated vielbein
(plus topological and Chern--Simons terms
in appropriate dimensions).

This result does not hold however, if the Yang--Mills
gauge group has abelian factors. Then new solutions involving
the conserved Noether currents (if any) and the antifields
exist and are discussed in the text.
\bsk

\noindent (iii) Candidate anomalies:

A similar result is derived for the anomalies.
Namely we show that the antifield dependence can be removed
from all candidate anomalies except if abelian factors
are present. Hence, for semi-simple Yang--Mills gauge groups
all candidate anomalies are given by the
well-known ``chiral anomalies" already discussed by
previous authors and by
``topological candidate
anomalies". The latter originate from the non-trivial De Rham
cohomology of the vielbein manifold and occur
in $n=n_0+4k$ dimensions where $n_0=6,9,20,35$ ($k=0,1,\ldots$).
All the chiral anomalies remain
non-trivial even on-shell unless the action contains
unconventional couplings.

In presence of abelian factors and
conserved Noether currents, there are further candidate anomalies
which are described in the text and involve the antifields.

\subsection{Outline of the paper}

The construction of the general solution of the
Wess--Zumino consistency condition \Gl{i1} is rather lengthy
and involves many technical steps. We therefore give in this
subsection a general outline of our strategy with an
emphasis on the main ideas. The details are developed in
the text.

There are in fact two essential ingredients in the calculation
of $H^{*,*}(s|d)$. The first one is the close relationship
that exists in diffeomorphism invariant theories between
$H^{*,*}(s|d)$ and the more standard BRST cohomological groups
$H^{*}(s)$. This was shown already
in \cite{grav,Brandt} where however the topological aspects of the
vielbein manifold have not been taken into account.
The cohomology groups $H^*(s)$ are defined as the set of equivalence
classes of solutions of the BRST cocycle condition
\beq s \a=0\label{i2}\eeq
modulo trivial solutions of the form $s\be$ where
$\a$ and $\be$ are local functions (0-forms).
It turns out that the knowledge of $H^{*}(s)$
enables one to completely construct $H^{*,*}(s|d)$.
The problem of computing the cohomology of $s$ modulo
$d$ can therefore be reduced to that of computing the
cohomology of $s$.  This property holds because the diffeomorphism
ghosts---which play somehow a role similar to the differentials
$dx^\m$---are included among the ghosts.
The relationship between $H^{*}(s)$ and $H^{*,*}(s|d)$ is derived
in sections \ref{basis}--\ref{3} where (i) a redefinition of
the dynamical variables is given that simplifies the form
of $s$ and the computation of its cohomology; (ii)
the cohomology of $d$ is investigated in the space of local
forms---it is here that the topological currents related to the
topology of the vielbein manifold appear; and (iii) the map
between $H^{*,*}(s|d)$ and $H^{*}(s)$ defined by the descent
equations is analyzed.

To find the general solution of the
Wess--Zumino consistency condition \Gl{i1}, it is thus
sufficient to compute $H^{*}(s)$. It is here that
the second important ingredient is used, namely
that $s$ is the sum of the so-called Koszul--Tate
differential and the ``longitudinal differential
along the gauge orbits". This feature is quite
standard and actually at the heart of the antifield
formalism \cite{henfisch}.
It enables us to compute $H^*(s)$ completely using
(a) the resolution properties of the Koszul--Tate complex,
(b) the known mathematical results on the Lie algebra cohomology
of the Lorentz and Yang--Mills gauge groups, and
(c) the equivariant characteristic cohomology which
we compute separately in appendix \ref{B} by means of the results
on the cohomology of the Koszul--Tate
differential modulo $d$ derived in \cite{bbh1}. This is the hard core
of an analysis carried out in section \ref{scoh}.

Sections \ref{ass}--\ref{discussion} are comparatively much
easier and simply put together the two ingredients
explained above to get the results announced in subsection \ref{1.3}.
We recommend these sections to readers who are interested in the
results but not in the details of the computation.
\newpage

\mysection{Assumptions and notation}\label{brs}

\noindent {\it Spacetime manifold:}

In principle, one should  study the general solution of the
Wess--Zumino consistency condition \Gl{i1} for gravity
coupled to
Yang--Mills gauge fields that may live on non-trivial
bundles.  Furthermore,
one should allow for general spacetime manifolds $M$.
This means that in general there is no
global section in the
bundle over $M$ of linear frames: it is not
possible to choose continuously
everywhere on spacetime a set of
$n$ linearly independent tangent vectors.
Similarly, the Yang--Mills connections
may not be globally defined (as Lie-algebra valued
$1$-forms over spacetime).

However,
before tackling the difficulties that arise because of
non-trivial global features, it appears necessary to
analyse the Wess--Zumino consistency condition
first on a local neighbourhood
$U$ homeomorphic to {\bf R}$^n$ which trivializes the bundle, i.e.
which is such that the bundle reduces,  over $U$, to the
trivial direct product $U\times F$ of $U$ with the
``fiber" $F$ of the field-antifield manifold. One can
then study how the local results can be pasted together
to provide global results. In this paper, we
completely solve the first step. Since the
neighbourhood $U$ is homeomorphic to {\bf R}$^n$, we shall actually
consider that $M$ is just {\bf R}$^n$. We carry out the
analysis for all $n>2$ (our methods apply also to the case $n=2$
but one encounters some subtleties in this case, cf. \cite{bbh1}).
\bsk

\noindent {\it Field content and gauge invariances:}

We consider gravitational theories whose classical action is invariant
under general coordinate transformations (diffeo\-mor\-phisms), local Lorentz
trans\-for\-mations and Yang--Mills gauge transformations. The Lorentz group
is denoted by $G_L$, the Yang--Mills gauge group by $G_{YM}$, their
Lie algebras by $\cG_L$ and $\cG_{YM}$ where the latter is supposed to be
reductive (= semi-simple + abelian).
We adopt the vielbein formulation
of gravity since we allow for matter fields
with integer and half-integer spins. The matter fields are denoted
by $y^\Ly$ and are assumed to transform scalarly under general coordinate
transformations (with no loss of generality, since world
indices can be converted to Lorentz indices by means of the
vielbein) and linearly under $G_{YM}\times G_L$.
The elements of $\cG_{YM}$ are
denoted by $\de_i$, the corresponding gauge fields
and ghosts by $\A \m{i}$ and $C^i$. The elements of $\cG_L$ are
denoted by $l_{ab}=-l_{ba}$,
the corresponding gauge fields (= components of the spin connection)
and ghosts by $\o \m{ab}=-\o \m{ba}$ and $C^{ab}=-C^{ba}$.
The independent elements of $\cG:=\cG_{YM}+\cG_L$ and the
corresponding gauge fields and ghosts are labeled by $I$:
\beann & & \{\de_I\}=\{\de_i,l_{ab}:\, a>b\} ,\\
 &&\{C^I\}=\{C^i,C^{ab}:\, a>b\},\qd
\{\A \m{I}\}=\{\A \m{i},\o \m{ab}:\, a>b\}.\eeann
The structure constants of $\cG$ and $\cG_{YM}$ are denoted by
$\f IJK$ and $\f ijk$ respectively:
\beann & &[\de_I,\de_J]=\f IJK\de_K\ ;\\ & &
[\de_i,\de_j]=\f ijk\de_k\ ,\qd [l_{ab},l_{cd}]=2\et_{a[c}l_{d]b}
-2\et_{b[c}l_{d]a}\ ,\qd [\de_i,l_{ab}]=0 \eeann
where $\et_{ab}$ are the entries of the
Minkowski metric (cf. appendix \ref{ap1}).

We define the spin connection
and the Christoffel connection through
\beq \6_\m\e \n{a}-\o {\m b}a\e \n{b}-\G \m\n\rho \e \rho{a}=0,
\qd \o \m{ab}=-\o \m{ba},\qd \G \m\n\rho=\G \n\m\rho
\label{0}\eeq
which determines $\o \m{ab}$ and $\G \m\n\rho$
in terms of the vielbein fields $\e \m{a}$ and their derivatives:
\bea \o \m{ab}&=&E^{a\n}E^{b\rho}(\om_{[\m\n]\rho}-\om_{[\n\rho]\m}+
\om_{[\rho\m]\n}),\qd \om_{[\m\n]\rho}=e_{\rho a}\6_{[\m}\e
{\n]}a,\label{0a}\\
\G \m\n\rho&=&\s0 12\, g^{\rho\si}(\6_\m g_{\n\si}+\6_\n g_{\m\si}-\6_\si
g_{\m\n}).\label{0b}\eea
Here $\E a\m$ are the components of the inverse vielbein and
$g_{\m\n},g^{\m\n}$ are the components of the spacetime metric
and its inverse
(Lorentz indices $a,b,\ldots$ are raised and lowered by means of the
Minkowski metric and its inverse):
\beq\E a\m\e \m{b}=\de_a^b\, ,\qd
\e \m{a}\E a\n=\de_\m^\n\, ,\qd  g_{\m\n}=e_{\m a}\e \n{a}\, ,\qd
g^{\m\n}=E^{a\m}\E a\n\, .\label{0c}\eeq
The set of fields is therefore given by
\[ \{\PH^A\}=\{\e \m{a},\A \m{i},y^\Ly,\xi^\m,C^I\}\]
where the $\xi^\m$ are the ghosts of diffeomorphisms.
The corresponding antifields are denoted by
\[ \{\PH^*_A\}=\{\hs a\m,\As i\m,y^*_\Ly,\xi^*_\m,C^*_I\}.\]

Note that \Gl{0a} and \Gl{0b} represent no loss of generality compared
with other choices of $\o \m{ab}$ and $\G \m\n\rho$, as considered
e.g. in \cite{schweda},
since non-vanishing torsion components
(or the fields they are built of)
can be counted among the matter fields $y^\Ly$.
\bsk

\noindent{\it Local functions and forms}:

We regard the
$\6_{\m_1\ldots\m_k}\PH^A$ and $\6_{\m_1\ldots\m_k}\PH^*_A$ ($k=0,1,\ldots$)
as local coordinates of a jet space $J^\infty$, cf.
\cite{bbh1} and references given there.
Throughout this paper a {\it local function} is by definition
a function of these generators
depending polynomially on all of them except on the
undifferentiated $\e \m{a}$ on which it must depend smoothly and regularly
for $\det (\e \m{a})\neq 0$. The algebra of local functions
is denoted by ${\cal A}$. A {\it local form} is by definition a
finite linear combination of local functions $\a_\tau$ with
coefficients $\om^\tau(x,dx)$ which are differential forms on
the spacetime manifold $M$ and whose algebra is
denoted by $\Om(M)$. The algebra of local forms is
denoted by ${\cal E}$. We thus have
\beann & &{\cal E}=\Om(M)\otimes {\cal A};\\
& & a\in {\cal E}\ \LRA\ a=
\sum_{\tau} \om^\tau\, \a_\tau,\qd \om^\tau\in\Om(M),
\ \a_\tau\in {\cal A}\eeann
where finiteness of the sum is understood.
Note that the explicit dependence on the coordinates and the form degree
of a local $p$-form are carried exclusively by $p$-forms
$\om^\tau\in\Om(M)$ whereas its dependence on the fields, antifields
and their derivatives and its ghost number is carried by local functions
$\a_\tau\in {\cal A}$.
\bsk

\noindent {\it Action functional and BRST operator:}

We assume that the classical action
\beq S_0=\int d^nx\, \cL_0(\phi,\6\phi,\ldots,\6^k\phi),\qd
\phi\equiv(\e \m{a},y^\Ly,\A \m{i})\label{xx0}\eeq
is an integrated local function $\cL_0$ and
does not possess any non-trivial gauge
invariances apart from those mentioned above (invariance under
diffeomorphisms and gauge invariance under $G_L\times G_{YM}$).
Hence, we exclude for instance
supergravity from the discussion. Then the
proper solution $\act$ of the classical master equation \cite{bv} is
given by
\bea \act=\int d^nx\lb\cL_0-\xi^\m\PH^*_A\6_\m\PH^A-C^Iy^*_\Ly\de_I y^\Ly
+\hs a\m(\e \n{a}\6_\m\xi^\n+\e \m{b}{C_b}^a)\right.\nn\\
\left.+\As i\m(\6_\m C^i+C^j\A \m{k}\f kji+\A \n{i}\6_\m\xi^\n)
+\s0 12 C^*_K C^IC^J\f IJK \rb.\label{xx6}\eea
Using the conventions of appendix \ref{ap1},
the BRST transformations of the fields and antifields are obtained
from $\act$ through
\beq s \PH^A(x)=-\0 {\de^R\act}{\de\PH^*_A(x)}\ ,\qd
s \PH^*_A(x)=\0 {\de^R \act}{\de\PH^A(x)}
\label{xx7}\eeq
where $\de^R/\de Z(x)$ denotes the
functional right derivative with respect to $Z$. The
BRST transformations of derivatives of the $\PH^A$ and $\PH^*_A$
and of local functions and forms are then
obtained from \Gl{xx7} using the rules
\beq s\6_\m-\6_\m s=0,\qd
s\, (XY)=(s\, X)Y+(-)^{\ep(X)}X(s\, Y)\label{xx8}\eeq
where $\ep(X)$ is the grading of $X$. The latter is defined through
\bea & & \ep(\e \m{a})=\ep(\A \m{i})=\ep(y^\Ly_B)=\ep(x^\m)=0,\nn\\
& & \ep(y^\Ly_F)=\ep(\xi^\m)=\ep(C^I)=\ep(dx^\m)=1,\nn\\
& &\ep(\PH^*_A)=\ep(\PH^A)+1\qd (\mbox{modulo}\ 2),\nn\\
& &\ep(\6_{\m_1\ldots\m_r}Z)=\ep(Z)\label{grad}\eea
where the $y^\Ly_B$ ($y^\Ly_F$) are the bosonic (fermionic) $y$'s.
The differentials and
coordinates have vanishing BRST transformation,
\beq  s x^\m=s dx^\m=0.\label{xx7a}\eeq

We shall carry out the analysis without assuming
a particular form of the classical action $S_0$, but we require that
it defines a normal theory in the terminology of \cite{bbh1}, apart
from the above-mentioned assumptions on the field content,
gauge invariances and locality. In order to use the definition
of \cite{bbh1} one has to split the vielbein into a constant background
vielbein and a deviation $h_\m^a$ from it, and linearize the action
resp. the field equations in the $h_\m^a$. We stress however that
we only need this split for defining the normality condition, but
not as a prerequisite for the computation.
In fact the normality of the theory
is a sufficient condition for the validity of our results,
not a necessary one, i.e. our results may apply
even to non-normal theories. We do not repeat
the definition of normal theories here, but we remark that most
theories of physical interest are normal.
An important example of a normal theory, which we will use
occasionally to illustrate the general results, is provided by
the usual Einstein--Yang--Mills--matter Lagrangian
\beq \cL_0/e= \s0 12\, R-\s0 14\, g^{\m\r}g^{\n\si}\F \m\n{i}F_{\r\si i}
+L^y(y^\Ly,D_a y^\Ly)\label{xx1}\eeq
where $L^y(y^\Ly,D_a y^\Ly)$ is at most linear (quadratic) in
the $D_a y^\Ly_F$ ($D_a y^\Ly_B$),
$e$ is the determinant of the vielbein,
$R=\R \m\n{ab}\E a\m\E b\n$ denotes the Riemann curvature
scalar, $\F \m\n{i}$
are the Yang--Mills field strengths\footnote{Indices $i,j,\ldots$
referring to the semi-simple part of $\cG_{YM}$
are raised and lowered by means of its Cartan-Killing metric;
indices $i,j,\ldots$ referring to abelian elements of $\cG_{YM}$ are
raised and lowered by means of the unit matrix.} and
$D_a y^\Ly$ are the covariant derivatives of the matter fields:
\bea e&=&\det (\e \m{a}),\label{xx2}\\
\R \m\n{ab}&=&\6_\m\o \n{ab}-\6_\n\o \m{ab}
+\o \m{ac}\o {\n c}b-\o \n{ac}\o {\m c}b\nn\\
&=&-E^{a\rho}\e \si{b}(\6_\m\G \n\rho\si-\6_\n\G \m\rho\si+
\G \m\la\si\G \n\rho\la-\G \n\la\si\G \m\rho\la),\label{xx3}\\
\F \m\n{i}&=&\6_\m\A \n{i}-\6_\n\A \m{i}+\f jki\A \m{j}\A \n{k},
\label{xx4}\\
D_a y^\Ly &=&\E a\m(\6_\m-\A \m{I}\de_I)y^\Ly .
\label{xx5}\eea

\mysection{New basis of generators of $\cA$}\label{basis}

Our fundamental variables are the fields and antifields and their
partial derivatives which according to section \ref{brs}
are considered as independent generators of the
algebra of local functions or,
in more mathematical terms,
as local coordinates of a jet space $J^\infty$ (together with
the $x^\m$). Following the lines of \cite{Brandt}, we shall however
carry out the analysis in terms of other generators which
isolates a manifestly contractible part of the algebra. The new
generators serve as local coordinates of $J^\infty$
as well, together with the $x^\m$.

The elements of the new basis that replace the fields $\PH^A$
and their partial derivatives can be grouped in four sets:
\bea & &\{\cT^r\}=\{D_{(a_1}\ldots D_{a_k)}y^\Ly,
D_{(a_1}\ldots D_{a_k}\F {a)}bI:\, k=0,1,\ldots\},
\label{b1}\\
& &\7\xi^a=\xi^\m\e \m{a},\qd
\7C^I=C^I+\xi^\m\A \m{I},
\label{b2}\\
& &\{U_l\}=\{\6_{(\m_1\ldots\m_k}\e {\m)}{a},
\6_{(\m_1\ldots\m_k}\A {\m)}{I}:\, k=0,1,\ldots \},
\label{b3}\\
& &\{V_l\}=\{\6_{(\m_1\ldots\m_k}s\e {\m)}{a},
\6_{(\m_1\ldots\m_k}s\A {\m)}{I}:\, k=0,1,\ldots\}
\label{b4}\eea
where
\beq \F abi=\E a\m\E b\n\F \m\n{i},\qd\F ab{cd}=
\E a\m\E b\n\R \m\n{cd},\qd D_a=\E a\m(\6_\m-\A \m{I}\de_I)\label{b1a}\eeq
and it is understood that only algebraically independent
curvature components $\F abI$ enter in \Gl{b1}
(together with the symmetrization of the indices in
\Gl{b1} this guarantees the absence of
algebraic identities between the generators \Gl{b1}, taking into
account $[D_a,D_b]=-\F abI\de_I$ and the resulting
Bianchi identities).
One easily verifies that indeed each local function of the
$\6_{\m_1\ldots\m_k}\PH^A$
can be expressed as a local function of the
generators \Gl{b1}--\Gl{b4} and vice versa\footnote{A local function
of the generators \Gl{b1}--\Gl{b4} depends polynomially on all of them
apart from $\e \m{a}$.}.
Skipping the details we just note that
(a) the change of generators $\7\xi^a,\7C^I\lra\xi^\m,C^I$
(given the $U_l$) is invertible; (b) the derivatives
of the ghosts can be replaced by the $V_l$ due to
$s\e \m{a}=\6_\m\7\xi^a+\ldots$, $s\A \m{I}=\6_\m C^I+\ldots$;
(c) the $D_{(a_1}\ldots D_{a_k)} y^\Ly$ are equivalent to the
$\6_{\m_1\ldots\m_k}y^\Ly$; (d) the $\o \m{ab}$ and
$\6_{(\m}\e {\n)}a$
are equivalent to the $\6_\m \e \n{a}$, cf. \Gl{0a}; (e) the
$\6_{(\m_1\ldots\m_{k+2}}\e {\m)}{a}$,
$\6_{(\m_1\ldots\m_{k+1}}\A {\m)}{I}$ and
$D_{(a_1}\ldots D_{a_k}\F {a)}bI$ are equivalent to
the $\6_{\m_1\ldots\m_{k+2}}\e {\m}{a}$ and
$\6_{\m_1\ldots\m_{k+1}}\A {\m}{I}$
(for each $k\geq 0$ separately).

The new generators that replace the undifferentiated
antifields are
\bea & &\7y^*_\Ly=y^*_\Ly /e,\qd
\7e^*_a{}^b=\e \m{b}\hs a\m/e,\qd
\7A_i^{*a}=\e \m{a} \As i\m/e,\nn\\
& & \7C^*_I=C^*_I/e,\qd
\7\xi^*_a=\E a\m(\xi^*_\m-\A \m{I}C^*_I)/e.
\label{b5}\eea
Apart from the more involved form of
$\7\xi^*_a$ these definitions just convert the original antifields
to generators transforming scalarly under general coordinate
transformations [the factor $1/e$ occurs because all antifields
transform as scalar or vector
densities under general coordinate
transformations as follows from \Gl{xx6} and \Gl{xx7}]. The new
generators
that replace the partial derivatives of the $\PH^*_A$ are
the corresponding symmetrized covariant derivatives of the
generators \Gl{b5}:
\beq  \{\cT^*_\oor\}=\{D_{(a_1}\ldots D_{a_k)}\7\PH^*_A:\, k=0,1\ldots\},
\qd \{\7\PH^*_A\}=\{\7y^*_\Ly,\7e^*_a{}^b,\7A_i^{*a},
\7C^*_I,\7\xi^*_a\}.\label{b6}\eeq

As mentioned above we shall compute the BRST cohomology using the new
generators. We shall therefore need their BRST transformations.
Those of the $U_l$ and $V_l$ are extremely
simple in the new basis since they just read
\beq sU_l=V_l, \qd sV_l=0.\label{trivpair}\eeq
The BRST transformations of
$\7\xi,\7C, \cT$ and $\cT^*$ are more complicated. For later purpose we
decompose the BRST operator according to
\beq s=\de+\g\label{b7}\eeq
where $\de$ is the Koszul--Tate differential
which has been discussed and analyzed in the antifield
context first in \cite{henfisch}. As we shall see it plays a crucial
role in the subsequent calculations. What distinguishes
$\de$ and $\g$ is
the antighost number ($antigh$) defined through
\beq\ba{l}\agh{\PH^A}=0,\\
\agh{y^*_\Ly}=\agh{e_a^{*\m}}=\agh{\As i\m}=1,\\
\agh{\xi^*_\m}=\agh{C^*_I}=2.\ea
\label{agh}\eeq
The ghost number ($gh$) is related to the antighost number by
\beq gh=puregh-antigh\label{ghostnumber}\eeq
where $puregh$ is defined through
\beq\ba{l}
puregh(\xi^\m)=puregh(C^I)=1,\\
puregh(y^\Ly)=puregh(\A \m{i})=puregh(\e \m{a})=puregh(\PH^*_A)=0.\ea\eeq
The differential
$\de$ is characterized by $\agh \de=-1$, i.e. it
lowers the antighost number by one unit and
acts on $\7\xi^a$, $\7C^I$, $\cT^r$, $\cT^*_\oor$ according to
\bea
& &\de\, \cT^r=\de\, \7C^I=\de\, \7\xi^a=0,\nn\\
& &\de\, \7y^*_\Ly=\0 1e\, \0 {\de\cL_0}{\de y^\Ly},\qd
   \de\, \7A_i^{*a}=\0 1e\,\e \m{a}\0 {\de \cL_0}{\de \A \m{i}},\qd
   \de\, \7e^*_a{}^b=\0 1e\,\e \m{b}\0 {\de \cL_0}{\de \e \m{a}},\nn\\
& &\de\, \7C^*_i=-D_a\7A_i^{*a}+\7y^*_\Ly\de_iy^\Ly,\qd
   \de\, \7C^*_{ab}=-2\7e^*_{[ab]}+\7y^*_\Ly l_{ab}y^\Ly,\nn\\
& &\de\, \7\xi^*_a=-D_b\7e^*_a{}^b-\7A_i^{*b}\F bai+\7y^*_\Ly D_a y^\Ly,\nn\\
& &\de\, D_{a_1}\ldots D_{a_k}\7\PH^*_A=D_{a_1}\ldots D_{a_k}\de\,
   \7\PH^*_A\label{b8}\eea
where $\de \cL_0/\de \phi$ denotes the Euler-Lagrange right-derivative of
$\cL_0$ with respect to $\phi$,
\beq \0 {\de \cL_0}{\de \phi}=\sum_{k\geq 0}(-)^k\,
\6_{\m_1}\ldots \6_{\m_k}
\0 {\6^R\cL_0}{\6(\6_{\m_1\ldots\m_k}\phi)}\ .
\label{euler}\eeq
We remark that $\de \7y^*_\Ly$, $\de \7A_i^{*a}$ and $\de \7e^*_a{}^b$
can be always expressed entirely in terms of the $\cT^r$, independently
of the particular choice of $\cL_0$,
since the equations of motion are necessarily covariant (in fact this
can be easily proved using the decomposition $\de+\gamma$ of $s$ and
its nilpotency). For instance, \Gl{xx1} yields
\bea
\de\, \7A_i^{*a}&\simple&
D_b F^{ba}{}_i+\0 1e\,\e \m{a}\0 {\de (eL^y)}{\de \A \m{i}},\nn\\
\de\, \7e^*_a{}^b&\simple&- {R_a}^b+ \s0 12\de_a^bR+\F aci F^{bc}{}_i
-\s0 14 \de_a^b\F dci F^{dc}{}_i
\nn\\  & &
+\0 1e\,\e \m{b}\0 {\de (eL^y)}{\de \e \m{a}}\label{b8simple}\eea
where $R_{ab}=R_{acb}{}^c$ is the Ricci tensor.

The differential
$\g$ is characterized by $\agh \g=0$ and
 acts on $\7\xi^a,\7C^I$, $\cT^r$, $\cT^*_\oor$ according to
\bea & &\g\, \cT^r=(\7\xi^a D_a+\7C^I\de_I)\cT^r,\qd
\g\, \cT^*_\oor=(\7\xi^a D_a+\7C^I\de_I)\cT^*_\oor,\nn\\
& &\g\, \7\xi^a=\7C_b{}^a\7\xi^b,\qd
\g \7C^I=\s0 12\, \7C^J\7C^K\f KJI+\7F^I\label{b9a}\eea
where
\beq \7F^i=\s0 12\, \7\xi^a\7\xi^b\F abi,\qd
\7F^{ab}=\s0 12\, \7\xi^c\7\xi^d\R cd{ab}.\label{b9b}\eeq
The nilpotency of $s$ and the different antighost numbers
of $\de$ and $\g$ imply
\beq \de^2=\g^2=\de\g+\g\de=0.\label{nilpot}\eeq

Using the new generators, the algebra of local
functions can be written as the tensor product of two algebras,
\beq {\cal A} = {\cal A}_1 \otimes {\cal A}_2, \label{tensorproduct1}\eeq
where ${\cal A}_1$ is generated by the $U_l$ and the $V_l$, and
${\cal A}_2$ is the algebra generated by
the tensor fields ${\cal T}^r, {\cal T}^*_\oor$ and the
undifferentiated ghosts $\hat \xi^a$ and $\hat C^I$. Note that,
by definition of $\cA$, ${\cal A}_2$ is a
polynomial algebra whereas ${\cal A}_1$ is polynomial in all
generators $U_l$ and $V_l$ but the undifferentiated vielbeins.
The usefulness of the new basis of generators
originates in the fact that both ${\cal A}_1$ and ${\cal A}_2$ are
$s$-invariant subalgebras of ${\cal A}$, i.e. the image of $\cA_i$
under $s$ is contained in $\cA_i$:
\beq s [{\cal A}_1]\subset {\cal A}_1,\qd
s [ {\cal A}_2]\subset {\cal A}_2.
\label{tensorproduct2}\eeq
This is evident from \Gl{trivpair}, \Gl{b8} and \Gl{b9a}, and
allows one to apply the K\"unneth formula
for the computation of the BRST cohomology in the ghost number
$g$ section of ${\cal A}$, denoted
by $H^g(s,\cA)$:
\beq s\a^g=0,\qd \a^g\in\cA\qd\LRA\qd \a^g=\sum_{k=0}^g
\a_{(1)}^k \a_{(2)}^{g-k}+s\be^{g-1},\qd \be^{g-1}\in\cA.\label{skunneth}\eeq
Here superscripts denote the ghost number and $\a_{(i)}^k$
denote representatives of cohomology classes of $H^k(s,\cA_i)$ for $i=1,2$.

\mysection{Cohomology of $d$ in the algebra ${\cal E}$ of local forms}
\label{APL}

Our first step is to characterize the cohomology of the
exterior derivative operator $d$ in the algebra of local forms.
As we shall see, this first
step is not devoid of global subtleties because the manifold
of the vielbeins carries some non-trivial cohomology. We
shall encounter topological currents of a type
reminiscent of the currents described in \cite{Fink}, which
must be taken into account for a correct analysis.

We denote by $E$ the bundle {\bf R}$^n\times F$ where, as we have already
mentioned, $F$ is the fiber of the fields and antifields.
Any point in $E$ can be parametrized by
$(x^\m$,\,$\PH^A$,\,$\PH^*_A)$.
These functions are all globally defined on
$E$.  However, they fail to provide global
free coordinates of $E$ because
the vielbein components $\e \m{a}$ are constrained
by the condition $\det (\e \m{a}) > 0$ (we assume the
vielbeins to have positive orientation), and so, do not have a
range of variation homeomorphic to $R^{n^2}$.  Rather, the
vielbeins belong to $GL^+(n)$, the set of matrices with
strictly positive determinant. Of course the range of values which
the matter fields can take may be restricted too,
but for simplicity we assume that the manifold of the
vielbeins is the only non-trivial factor of $E$,
so that $E$ can be contracted to it. Thus, effectively
one has $H^*_{DR}(E) = H^*_{DR}(GL^+(n))$ which is non-trivial.
In the terminology of \cite{Fink}, the
manifold of the vielbeins is a ``non-linear" manifold.

The cohomology of the
exterior derivative operator $d$ in the algebra
${\cal E}$ of local forms is described by a theorem
which has been obtained in
\cite{Takens} and is discussed in \cite{Anderson}.  It
generalizes well known results derived and rederived for trivial bundles
by various people to a bundle $E$ which is not cohomologically
trivial (besides the above mentioned
references, see \cite{various,com}).
The theorem in question is in fact valid for bundles that
need not be direct products and
states explicitly the following:
\begin{theorem}{\em{\bf :}}\label{Poincare}
The cohomology $H^k(d,{\cal E})$ of $d$ in the algebra
of local $k$-forms is isomorphic to the De Rham cohomology of
$E$ in the same degree for $k<n$,
\begin{eqnarray}
H^k(d,{\cal E})\simeq H^k_{DR}(E), \;  k<n. \label{4.1}
\end{eqnarray}
Furthermore, in form degree $n$, one has
\begin{eqnarray}
\frac{\{\hbox{variationally closed $n$-forms}\}}
{\{d \hbox{-exact $n$-forms}\}} \simeq H^n_{DR}(E) \label{4.2}
\end{eqnarray}
where the $n$-form $\omega = {\cal L} d^nx$
is said to be ``variationally closed" iff the Euler-Lagrange
derivatives of ${\cal L}$ with respect to all the fields
and antifields identically vanish.
\end{theorem}

We shall not prove the theorem (see original literature).
We shall instead draw some of its consequences for the
theory at hand.

As we have pointed out, the De Rham cohomology of $E$ is
isomorphic to the De Rham cohomology of the fibers (the base
is trivial) and reduces to $H^*_{DR}(GL^+(n))$ since all factors
but the vielbein one are (effectively) trivial
(contractible to a point).
Now, the manifold $GL^+(n)$ is diffeomorphic (as a smooth manifold,
not as a group!) to the product manifold $N \times SO(n)$, where
$N$ is the set of upper triangular matrices with strictly
positive real numbers on the diagonal (Gram--Schmidt
decomposition).  The manifold $N$ is contractible.
Thus $SO(n)$ is a deformation retract
of $GL^+(n)$.  This implies that the De Rham cohomology of
$GL^+(n)$ is isomorphic to the De Rham cohomology of $SO(n)$,
\begin{eqnarray}
H^*_{DR}(GL^+(n))\simeq H^*_{DR}(SO(n)), \label{result}
\end{eqnarray}
and can be described in terms of the Lie algebra
cohomology of $so(n)$.

Given a non-trivial closed $k$-form $\alpha (\e \m{a},
d_{GL^+(n)} \e \m{a})$ on $GL^+(n)$ ($k\leq n$)\footnote{At
any point of $GL^+(n)$, the forms $d_{GL^+(n)} \e \m{a}$ are
well-defined and yield a basis of the cotangent space.  Thus,
any $k$-form on $GL^+(n)$ can be expressed as a polynomial of
order $k$ in the  $d_{GL^+(n)} \e \m{a}$ with coefficients
that are smooth functions of the $\e \m{a}$.}, one gets the
corresponding class $[\alpha]$ in $H^k(d,{\cal E})$ by
simply replacing the exterior derivative $d_{GL^+(n)}$
on $GL^+(n)$ by the spacetime exterior derivative $d$.

We shall denote the result by $\alpha^{0,k}$ to emphasize
the fact that one gets a $k$-form on ${\cal E}$ with
ghost number $0$.  Thus,
\begin{eqnarray}
\alpha^{0,k}(\e \m{a}, d \e \m{a}) =
\alpha (\e \m{a},
d_{GL^+(n)} \e \m{a} \rightarrow d \e \m{a})  \label{4.4}
\end{eqnarray}
with $d\e \m{a} \equiv dx^\rho\partial_\rho \e \m{a}$.  The Lie algebra
cohomology of $so(n)$ is finite-di\-men\-sional.
Thus, $H^k(d,{\cal E})\simeq H^k_{DR}(GL^+(n))\simeq H^k_{DR}(SO(n))$
is also finite-dimensional for $k<n$.  Let $\{[\alpha_m]\}$ be a basis
of the cohomology $H^*_{DR}(GL^+(n))$  and let
$\alpha_m^{0,k_m}(\e \m{a}, d \e \m{a})$ be the corresponding
local forms of ${\cal E}$ obtained through (\ref{4.4}).
Let $a^k$ be a closed $k$-form of ${\cal E}$ with $k<n$.
What the above theorem says is that one can express $a^k$ as
a linear combination of the $\alpha_m^{0,k_m}(\e \m{a}, d \e \m{a})$
of same degree modulo a $d$-exact form,
\begin{eqnarray}
k<n:\qd d a^k = 0 \qd\Leftrightarrow \qd a^k =
\sum_{\{m|k_m = k\}} \lambda_m
\alpha_m^{0,k_m} + d b^{k-1}, \label{4.5}
\end{eqnarray}
for some constants $\lambda_m$ and some ($k-1$)-form $b^{k-1}$.
Similarly, in degree $n$, one has
\begin{eqnarray}
\frac {\delta {\cal L}}{\delta Z} = 0\qd \forall Z\in\{\PH^A,\PH^*_A\}
\qd\Leftrightarrow\qd {\cal L} d^nx=
\sum_{\{m|k_m = n\}}
\lambda_m\alpha_m^{0,k_m} + d b^{n-1} \label{4.6}
\end{eqnarray}
for some constants $\lambda_m$ and some ($n-1$)-form $b^{n-1}$.
The coefficients $\lambda_m$ in the decompositions
(\ref{4.5}) and (\ref{4.6}) are unique, since the relation
$\sum \lambda_m
\alpha_m^{0,k_m} + d b = 0$ implies $\lambda_m = 0$.

To give an example of (\ref{4.6}), consider $3$-dimensional
gravity.  The $3$-form
\begin{eqnarray}
\alpha^{0,3} = \Om_a{}^b \Om_b{}^c \Om_c{}^a
\qd\mbox{with}\qd
\Om_a{}^b=\E a\m d\e \m{b}
\end{eqnarray}
is easily seen to be variationally closed, but
it cannot be written as the exterior derivative of a
globally defined $2$-form. The integral of
$\a^{0,3}$ is equal to the winding number of the map
from the spacetime manifold to the group manifold
defined by $\e \m{a}$. As $\a^{0,3}$ above, one
obtains in $n=2r+1$ space-time dimensions all $\a^{0,k_m}_m$ simply
through the substitution ${C_a}^b\rightarrow {\Om_a}^b$
from the corresponding ghost polynomials
$P(\Th(C))$ describing the Lie algebra cohomology
of $so(n)$, cf. appendix \ref{ap2}. Note that
$\Om^{ab}$ has both a symmetric and an antisymmetric part,
contrary to $C^{ab}=-C^{ba}$. One must not omit the symmetric
part of $\Om^{ab}$ since otherwise the resulting polynomials in
$\Om^{ab}$ would not be $d$-closed. This reflects that
$\a^{0,k_m}_m$ arises from a cohomology class of
$H_{DR}(GL^+(n))$ rather than of just $H_{DR}(SO(n))$. In $n=2r$
spacetime dimensions the situation is slightly more complicated.
Here one obtains all $\a^{0,k_m}_m$ as described above through
the substitution ${C_a}^b\rightarrow {\Om_a}^b$ from
the $P(\Th(C))$ except for
the one which corresponds to the primitive element
$\Th_r(C)$ in the notation of appendix \ref{ap2}
($\Th_r(\Om)$ is not $d$-closed). Since
$\Th_r(C)$ has ghost number $2r-1=n-1$, the corresponding
$\a^{0,k_m}_m$ is an $(n-1)$-form. It should therefore
agree with the $(n-1)$-form discussed in \cite{Torre} since further
$\a^{0,k_m}_m$ with form-degree $(n-1)$ do not exist in all
even spacetime dimensions
 but just for $n=4k$ and $n=22+4k$ ($k=0,1,\ldots$),
and since the other $\a^{0,k_m}_m$ are not gauge invariant under
local Lorentz rotations of the vielbein, cf. remark below.

In four dimensions, one gets non-trivial cohomology in
form degrees $0$ and $3$, with one cohomology class in degree 0
(represented by a constant) and two cohomology classes in degree 3.

Since the cohomology is completely carried by the vielbeins,
one has also the following immediate consequence of
Theorem \ref{Poincare}:
\begin{corollary}{\em{\bf :}}
Let $a^k$ be a
closed form of degree $k<n$ that vanishes when one sets
all the fields, the antifields and their derivatives equal
to zero, except the vielbeins and their derivatives.  Then,
$a^k$ is exact, $a^k = d b^{k-1}$.  Similarly,
let $a^n$ be a variationally closed
$n$-form that vanishes under the same conditions.
Then, $a^n$ is exact, $a^n = d b^{n-1}$.  In particular, if
$a^k$ is a (variationally) closed form of non-vanishing
ghost number, then $a^k = d b^{k-1}$.
\label{cor2}\end{corollary}

\noindent {\it Remark:}

We shall see in the next section that the forms $\alpha_m^{0,k_m}$
have counterparts $\alpha_m^{g,k_m-g}$ with non-vanishing ghost numbers
$g=1,\ldots,k_m$, arising even for $k_m>n$. Depending on $n$,
some of these forms contribute to solutions $a^{g,p}$ of \Gl{i1}
in physically interesting cases such as $(g,p)=(0,n)$ (possible
contributions to the action) or $(g,p)=(1,n)$ (candidate anomalies).
The $\a^{0,n}_m$ are somewhat analogous to the $\Theta$-term in
Yang--Mills theory. They are present
in $n=n_0+4k$ dimensions where $n_0=3,10,21,36$ ($k=0,1,\ldots$).

The physical interpretation of the $\a^{0,k_m}_m$ is obscured by the
fact that most of them have no counterpart in the metric formulation
of gravity. Namely in the metric formulation closed and
non-exact forms analogous to the $\alpha_m^{0,k_m}$ exist
only in even spacetime dimensions and only in form degree $(n-1)$,
if the metric has Lorentzian signature $(+,-,\ldots,-)$
\cite{Torre}.

Albeit the physical relevance or interpretation of the
different results in the metric and the vielbein formulation of gravity
remains to be uncovered, their mathematical origin is easily understood.
Namely, since the metric remains unchanged under local Lorentz rotations
of the vielbein $\e \m{a}(x)\rightarrow
\e \m{b}(x)\Lambda_b{}^a(x)$ ($\Lambda\in SO(1,n-1)$),
only those linear combinations of the
$\alpha_m^{0,k_m}$ can have counterparts in the metric formulation
which can be cast in a form (by adding exact forms, if necessary)
that is gauge invariant under
Lorentz transformations. This invariance condition eliminates
a great number of cohomological classes and leaves one with
the De Rham cohomology of the (Lorentzian) metric manifold,
which is two (one) dimensional in even (odd)
dimensional spacetime (see e.g. \cite{Torre}).

\mysection{BRST cohomology in $\cA_1$ and
K\"unneth formula for $H^*(s,{\cal E})$}\label{2}

The De Rham cohomology of $GL^+(n)$ appears again when one
analyzes the BRST cohomology in
the algebra ${\cal E}=\Omega(M) \otimes {\cal A}$ of local forms.
We have seen in section \ref{basis} that the algebra ${\cal A}$
of local functions is the direct product of two $s$-invariant
subalgebras ${\cal A}_1$ and ${\cal A}_2$ generated by
$\{U_l,V_l\}$ and $\{{\cal T}^r, {\cal T}^*_\oor,
\hat \xi^a,\hat C^I\}$ respectively, cf.
\Gl{tensorproduct1} and \Gl{tensorproduct2}. The cohomology
$H^*(s,{\cal A}_2)$ of $s$ in ${\cal A}_2$ will be
discussed at length in the sequel.  We compute here
$H^*(s,{\cal A}_1)$ and extend the K\"unneth formula \Gl{skunneth}
for $\cA$ to ${\cal E}$.

\begin{theorem}{\em{\bf :}}\label{5.1}
The cohomology $H^*(s,{\cal A}_1)$ of $s$ in
${\cal A}_1$ is finite-dimensional and
isomorphic to the De Rham cohomology
of $GL^+(n)$,
\begin{eqnarray}
H^*(s,{\cal A}_1)\simeq H^*_{DR}(GL^+(n)). \label{s_in_a1}
\end{eqnarray}
\end{theorem}

\proof{The cohomology of $s$ in ${\cal A}_1$ would be
trivial if the vielbeins were not restricted to belong to
 $GL^+(n)$. Indeed, $s$ takes in ${\cal A}_1$ the standard
contractible form due to \Gl{trivpair}. Therefore the
generators ($U_l, V_l$) disappear
in pair whenever $U_l$ ranges over the entire real line.
This leaves one with the single pairs ($\e \m{a},
s\e \m{a}$), for which
there is the constraint $\det (\e \m{a}) >0$.

Now, the algebra of polynomials in $s\e \m{a}$ with coefficients
that are smooth functions of $\e \m{a}$ (on $GL^+(n)$) is
isomorphic to the ex\-ter\-ior al\-ge\-bra $\Omega(GL^+(n))$, and the
actions of $s$ and $d_{GL^+(n)}$ coincide in the
isomorphism.  Thus
$H^*(s,{\cal A}_1)$ $\simeq  H^*_{DR}(GL^+(n))$, which
proves the theorem.}

We shall denote by $\alpha^{k_m,0}_m(\e \m{a}, s\e \m{a})$ the
elements in ${\cal A}_1$ corresponding to the representatives of
the classes of $H^*_{DR}(GL^+(n))$ introduced above.  These
forms are simply obtained from $\alpha_m$ by replacing
$d_{GL^+(n)}$ by $s$ (rather than by $d$).  They have
ghost number $k_m$ and form degree zero.

For $k_m\leq n$
one may relate very simply $\alpha^{k_m,0}_m(\e \m{a}, s\e \m{a})$
to  $\alpha_m^{0,k_m}(\e \m{a}, d \e \m{a})$ through descent
equations.  Let
\begin{eqnarray}
\tilde s = s + d
\end{eqnarray}
and let $\tilde \alpha_m =
\alpha_m(\e \m{a}, d_{GL^+(n)}\e \m{a} \rightarrow \tilde s \e \m{a}$).
$d_{GL^+(n)} \alpha_m = 0$ implies
$\tilde s \tilde \alpha_m = 0$.  If one expands
this equation according to the ghost number, one gets a chain of
``descent equations" that read explicitly
\beq\ba{c}
d \alpha^{0,k_m}_m = 0,\\
s\alpha^{0,k_m}_m + d \alpha^{1,k_m-1}_m = 0,\\
 \vdots \\
s\alpha^{k_m-1,1}_m + d \alpha^{k_m,0}_m = 0,\\
s\alpha^{k_m,0}_m = 0.
\ea\label{alphadesc}\eeq
Because $\alpha^{0,k_m}_m$ defines a non-trivial class in
$H^{0,k_m}(d,{\cal E})$,
the first of these equations does not
imply that $\alpha^{0,k_m}_m$ is $d$-exact ($k_m\leq n$).  This is
of course necessary for the consistency of the analysis,
since $d$-triviality of the top form $\alpha^{0,k_m}_m$ would
imply $s$-triviality of the bottom form $\alpha^{k_m,0}_m$.

Actually, one can even prove more about $\alpha^{0,k_m}_m$.
Namely, that it is a non-trivial element of $H^{0,k_m}(s|d)$ (i.e.,
that it remains non-trivial even on-shell).  Indeed, if one
had $\alpha^{0,k_m}_m = d \beta^{0,k_m-1}_m + s
\beta^{-1,k_m}_m$, then one would get from
$s\alpha^{0,k_m}_m + d \alpha^{1,k_m-1}_m = 0$ that
$\alpha^{1,k_m-1}_m - s \beta^{0,k_m-1}_m$ is $d$-closed.  But
since the latter has ghost number equal to one, it is exact by
corollary \ref{cor2}.  This
implies that $\alpha^{1,k_m-1}_m$ is also trivial in $H^{1,k_m-1}(s|d)$,
$\alpha^{1,k_m-1}_m = d \beta^{1,k_m-2}_m + s \beta^{0,k_m-1}_m$.
Repeating the argument would finally lead to the conclusion
that the last term $\alpha^{k_m,0}_m$ is also trivial in
$H^{k_m}(s)$, i.e.  $\alpha^{k_m,0}_m = s\beta^{k_m-1,0}_m$ (there is no
($-1$)-form). This is the desired contradiction
since $\alpha^{k_m,0}_m$ is not $s$-exact.
The same argument applies to linear combination of the
$\alpha^{0,k_m}_m$'s.  Thus we have :

\begin{theorem}{\em{\bf :}}\label{5.2} No non-vanishing linear
combination of the $\alpha^{0,k_m}_m$ is trivial in
$H^{*,*}(s|d)$,
\begin{eqnarray}
\sum_{\{m|k_m\leq n\}} \lambda_m \alpha^{0,k_m}_m = s b + d c  \qd\LRA\qd
\lambda_m = 0\qd \forall\, m.
\end{eqnarray}
\end{theorem}

Of course, this property is shared by all non-vanishing
$\alpha^{k_m -p,p}$, even for
$k_m>n$, as the same argument shows (if $k_m>n$ the
descent equations \Gl{alphadesc} remain valid with
$\alpha^{k_m-p,p}_m\equiv 0$ for $p>n$).
Finally, we leave it to the reader to check that the following
theorem holds:
\begin{theorem}{\em{\bf :}}\label{5.3}
The cohomology $H^*(s,{\cal E})$ of $s$ in
the algebra ${\cal E} = \Omega(M) \otimes {\cal A}_1
\otimes {\cal A}_2$ is given by
$\Omega(M) \otimes H^*(s, {\cal A}_1)
\otimes H^*(s, {\cal A}_2)$.  That is, if $a$
is BRST closed, $s a = 0$, then
\begin{eqnarray}
a = \sum_{m,A} \om_{m,A}(x,dx) \,\alpha^{k_m,0}_m
(\e \m{a}, s\e \m{a}) \, P_A({\cal T,T}^*,\hat \xi, \hat C)
+ s b, \label{prod1}
\end{eqnarray}
where the polynomials $P_A$ define a basis of
the cohomology $H^*(s,{\cal A}_2)$ (to
be computed in section \ref{scoh}) and
where the $\om_{m,A}(x,dx)$ are spacetime forms.  Furthermore,
if
\begin{eqnarray}
\sum_{m,A} \om_{m,A}(x,dx) \,\alpha^{k_m,0}_m
(\e \m{a}, s\e \m{a}) \, P_A({\cal T,T}^*,\hat \xi, \hat C)
= s b  \label{prod2}
\end{eqnarray}
holds for some local form $b$,
then all the form-coefficients $\om_{m,A}(x,dx)$ vanish.
[Finiteness of the sums in (\ref{prod1}) and
(\ref{prod2}) is understood since the forms are local by assumption].
\end{theorem}

\mysection{Descent equations}\label{3}

The relationship between $\alpha^{0,k_m}_m$ and
$\alpha^{k_m,0}_m$ found above provides a good example of the descent
equations techniques \cite{StoraZumino}. These play a crucial
role in our investigation of $H^{*,*}(s|d)$ and are developed in detail
in this section.
Let $a^{k,q}$ be a local $q$-form of ghost number $k$.
Assume that $a^{k,q}$ is a non-trivial solution
of the Wess--Zumino consistency condition,
\begin{eqnarray}
sa^{k,q} + d a^{k+1,q-1} = 0,  \label{WZ}
\end{eqnarray}
\begin{eqnarray}
a^{k,q}  \not= sb^{k-1,q} + db^{k,q-1}.
\end{eqnarray}
By applying $s$ to (\ref{WZ}), one gets $ d( s a^{k+1,q-1})
=0$.  Thus, $s a^{k+1,q-1}$ is a closed local form.
It follows from (\ref{4.5}) that
$s a^{k+1,q-1} + d a^{k+2,q-2} = \sum \lambda_m
\alpha^{0,k_m}_m$, where the $\lambda_m$ can be
non-vanishing only if $k+2 = 0$.  But even in that case, theorem
\ref{5.2} implies $\lambda_m = 0$.  Thus, $a^{k+1,q-1}$
is itself also a solution of the Wess--Zumino
consistency condition,
\begin{eqnarray}
s a^{k+1,q-1} + d a^{k+2,q-2} = 0. \label{WZbis}
\end{eqnarray}
It should be noted that $a^{k+1,q-1}$ is not completely determined
by (\ref{WZ}), even as a representative of $H^{k+1,q-1}(s|d)$.
Indeed, $a^{k,q}$
itself is determined only up to the addition of
trivial solutions, i.e., up to the addition of
a $s$-coboundary modulo $d$.  The resulting ambiguity in
$a^{k+1,q-1}$ reads
\begin{eqnarray}
a^{k+1,q-1} \rightarrow a^{k+1,q-1} + s b^{k,q-1} +
d b^{k+1,q-2} + \sum \lambda_m \alpha^{0,k_m}_m
\label{ambiguity}
\end{eqnarray}
since the last term in (\ref{ambiguity}) disappears
from (\ref{WZ}).  The presence of the $\alpha^{0,k_m}_m$'s
is permitted for $k=-1$ and plays no incidence
on the analysis of the descent equations.

Now, two things can happen :

\noindent
(i) either $a^{k+1,q-1}$ is trivial in $H^{k+1,q-1}(s|d)$ modulo
a combination of the $\alpha^{0,k_m}_m$'s; in this case,
one may eliminate it by adding to $a^{k,q}$ a $d$-exact term;
once this is done, $a^{k,q}$ is
$s$-closed, $sa^{k,q} = 0$; one says that $a^{k,q}$ generates
a trivial descent;

\noindent
(ii) or  $a^{k+1,q-1}$ is non-trivial in $H^{k+1,q-1}(s|d)$,
even up to a combination of the $\alpha^{0,k_m}_m$'s;
in this second case, there is no way to make $a^{k,q}$
$s$-closed by adding to it trivial solutions;
we shall
show that the descent then keeps going all the
way down to the zero forms, where it
hits an element of $H^*(s,{\cal A})$.
One says that $a^{k,q}$ generates
a non-trivial descent.

The argument for establishing this fundamental
property of non-trivial descents
proceeds along lines extremely close to those of
\cite{grav,Brandt}, where, however, only the case $q=n$ is discussed
and the extra solutions arising from the cohomology of the
vielbein manifold are not considered.
It goes as follows.  If one acts with
$s$ on (\ref{WZbis}) and uses again (\ref{4.5}) and
theorem \ref{5.2}, one finds that $a^{k+2,q-2}$ is also
a solution of the Wess--Zumino consistency condition.
The procedure can be repeated until  one hits a form
$a^{k+u,q-u}$ that is $s$-closed,
\begin{eqnarray}
sa^{k+u,q-u} = 0 , \qd u \leq q,  \label{sclosed}
\end{eqnarray}
which one eventually does (at the latest for $u=q$ since the Wess--Zumino
condition reduces to (\ref{sclosed}) for $0$-forms).
If $a^{k+u,q-u}$ is trivial in $H^{k+u,q-u}(s|d)$ modulo
a combination of the $\alpha^{0,k_m}_m$'s, then one may
assume $a^{k+u,q-u} = 0$ by
adding to $a^{k+u-1,q-u+1}$
a $d$-closed term.  The new $a^{k+u-1,q-u+1}$
is clearly  $s$-closed,
$sa^{k+u-1,q-u+1} = 0$.  The redefinition
 does not modify the previous
$a^{i,j}$'s and shortens the descent by one step.  We
shall assume that such redefinitions have already
been made.  Therefore,  the
last form $a^{k+u,q-u}$ in the descent is non-trivial in
$H^{k+u,q-u}(s|d)$, even modulo combinations of the
$\alpha^{0,k_m}_m$'s, and thus certainly non-trivial
in $H^*(s,{\cal E})$.  By assumption, we have  also
$u>0$ (non-trivial descent).

According to theorem \ref{5.3}, $a^{k+u,q-u} $ takes the form
\begin{eqnarray}
a^{k+u,q-u} = \sum_{m,A} \om_{m,A}(x,dx) \,\alpha^{k_m,0}_m
(\e \m{a}, s\e \m{a}) \, P_A({\cal T,T}^*,\hat \xi, \hat C),
\label{expansion}
\end{eqnarray}
modulo $s$-trivial terms that can be absorbed
through redefinitions.  The crucial
observation at this point is that all the
elements $\alpha^{k_m,0}_m
\, P_A$ of $H^*(s,{\cal A})$
also come from a non-trivial descent, i.e.,
their $d$ variation is $s$-exact.
This was seen
to be true for $\alpha^{k_m,0}_m$ already in the previous
section but holds in fact for any BRST invariant local function
due to a general property of gravitational
theories. Namely the presence of
diffeomorphisms among the gauge transformations implies that the
derivative operator $\6_\m$ can be represented on
the fields, antifields and their
derivatives by $\6_\m=sb_\m+b_\m s$ with $b_\m=\6/\6\xi^\m$.
Therefore the exterior derivative of a local function
can be represented by a commutator:
\beq \a\in{\cal A}\qd\then\qd d \a=[b,s] \a\label{drepres1}\eeq
where $b$ is the operator
\beq b=dx^\m\, \0 \6{\6 \xi^\m}\ .\label{drepres2}\eeq
Of course it is understood in \Gl{drepres1} that $\a$ is expressed
in terms of the orginial ghosts $\xi^\m$ and $C^I$ rather than in terms
of the $\7\xi^a$ and $\7C^I$. In particular \Gl{drepres1} implies
\begin{eqnarray}
d (\alpha^{k_m,0}_m\, P_A) = s \sigma_{mA},\qd
\sigma_{mA}=-b(\alpha^{k_m,0}_m\, P_A) \label{lift}
\end{eqnarray}
and thus
\begin{eqnarray}
d a^{k+u,q-u} = s \sum_{m,A} \om_{m,A}(x,dx) \,
\sigma_{mA} + \sum_{m,A} (d\om_{m,A}(x,dx)) \, \alpha^{k_m,0}_m
\, P_A.
\end{eqnarray}
But $d a^{k+u,q-u}$ is $s$-exact since, by assumption, $a^{k+u,q-u}$
comes from a non-trivial descent. Accordingly, we get
\begin{eqnarray}
\sum_{m,A} (d\om_{m,A}(x,dx)) \alpha^{k_m,0}_m
\, P_A = s (\hbox{something}),
\end{eqnarray}
which by theorem \ref{5.3} is possible only if
$d\om_{m,A}$ vanishes. Thus the coefficient forms
$\om_{m,A}$ of $\alpha^{k_m,0}_m\, P_A$ in (\ref{expansion})
are closed forms. Hence,
if $q-u>0$, the forms $\om_{m,A}$ are exact (Poincar\'e lemma
on {\bf R}$^n$). But then one gets, using (\ref{expansion})
and (\ref{lift}), that $ a^{k+u,q-u}$ is trivial in $H^{k+u,q-u}(s|d)$,
contrary to the assumption.  Thus $q=u$, and the coefficients
$\om_{m,A}$ in the expansion (\ref{expansion}) are constants,
\begin{eqnarray}
a^{k+q,0} = \sum_{m,A} \la_{m,A}\,\alpha^{k_m,0}_m
(\e \m{a}, s\e \m{a}) \, P_A({\cal T,T}^*,\hat \xi, \hat C),\qd
\la_{m,A}=const.\ .
\label{expansion2}
\end{eqnarray}
This indicates that indeed any non-trivial descent ends
on a zero-form in the algebra ${\cal A}$, which is a
non-trivial $s$-cocycle. Conversely there are no
obstructions to lift any $s$-cocycle $\a^g\in{\cal A}$
to solutions $a^{g-q,q}$ of \Gl{WZ} given by
\beq a^{g-q,q}=\0 1{q!}\, b^q\, \a^g\label{reconstr}\eeq
due to \Gl{drepres1}. Thus the ``integration" of the
descent equations is trivial. Note however that the
tower of forms obtained from \Gl{reconstr}
terminates at some form degree $q=q_{max}$
which may be smaller than the spacetime dimension
(viewing $\a^g$ as a polynomial in the
undifferentiated $\xi^\mu$, $q_{max}$ is
the degree of that polynomial).
In particular one has $da^{g-q_{max},q_{max}}=0$. For $q_{max}<n$
we conclude, using \Gl{4.5}, that $a^{g-q_{max},q_{max}}$
is equivalent to a linear combination of the $\a^{0,k_m}_m$ with
$k_m=q_{max}$, and thus that $\a^g$ is equivalent to a
linear combination of the corresponding $\a^{k_m,0}_m$,
cf. \Gl{alphadesc}.
Furthermore, arguments used already within
the derivation of theorem \ref{5.2} show that the $a^{g-q,q}$
($q=0,\ldots,q_{max}$) obtained from \Gl{reconstr} solve
\Gl{WZ} non-trivially if $\a^g$ is a non-trivial $s$-cocycle
(and vice versa). This proves

\begin{theorem}\label{correspondence}
Any solution of \Gl{i1} generating a non-trivial descent
corresponds via the descent equations to a non-trivial $s$-cocycle
$\a\in{\cal A}$ and can be obtained from the latter through
\Gl{reconstr}. Conversely, any non-trivial $s$-cocycle
$\a\in{\cal A}$ gives rise to a non-trivial solution of the
descent equations given by the tower of
forms obtained from $\a$ through
\Gl{reconstr}. This tower of forms ranges always from a 0-form to
an $n$-form unless $\a$ is equivalent to a
linear combination of the $\a^{k_m,0}_m$ with $k_m<n$ (including
the constant).
\end{theorem}
This result implies immediately that a
BRST invariant function $\a\in\cA_2$ with ghost
number $<n$ is BRST exact up to a constant
unless it depends
non-trivially on antifields. Indeed, suppose that $\a$ is
non-trivial. Then it corresponds by theorem \ref{correspondence}
to a non-trivial volume form or it is equivalent to a
constant. The volume form has negative
ghost number due to $\gh{\a}<n$ and therefore
must depend on antifields.
Since the ``ascent procedure" \Gl{reconstr} does not introduce
any antifield dependence, we conclude that $\a$ involves
also antifields or is $s$-exact up to constant.
\begin{corollary}
Any antifield independent BRST invariant function $\a\in\cA_2$
with ghost number $G<n$ is BRST exact (for $G\neq 0$) or
equivalent to a constant (for $G=0$),
\bea & &s\a=0,\qd \a\in\cA_2,\qd \gh{\a}=G<n,\qd \agh{\a}=0
\nn\\
& &\then\qd \a=s\be+\la \de_G^0,\qd \la=const.\ .\eea
\label{cor1}
\end{corollary}

So far we have only treated the case in which
$a^{k,q}$ generates a non-trivial descent. The case of a trivial
descent is easily discussed since $sa^{k,q}=0$ implies
according to theorem \ref{5.3} that $a^{k,q}$ is of the
form \Gl{prod1}. We note that the coefficients $\om_{m,A}$
occurring there can be assumed to be non-exact forms since
a contribution $\om_{m,A}\alpha^{k_m,0}_m P_A$ can be
removed from $a^{k,q}$ through a redefinition $a^{k,q}$
$\rightarrow$ $a^{k,q}+sb^{k-1,q}+db^{k,q-1}$ whenever
$\om_{m,A}$ is exact as a consequence of \Gl{lift}. In particular,
non-trivial volume forms $a^{k,q}$ have always a non-trivial descent
since all volume forms $\om_{m,A}(x,dx)$ are exact (Poincar\'e lemma
in {\bf R}$^n$). We thus obtain
\begin{theorem}\label{trivialdesc}
Any non-trivial solution of \Gl{i1} generating a trivial descent
has the form \Gl{prod1} where all coefficient forms $\om_{m,A}(x,dx)$
can be assumed to be non-exact. In particular, non-trivial
volume forms generate always a non-trivial descent.
\end{theorem}

Note that, due to the absence of non-trivial volume forms generating
a trivial descent,
$H^{g,n}(s|d)$ is isomorphic to $H^{g+n}(s,\cA)$ after dividing
out the constants,
\beq H^{g-n,n}(s|d)\simeq H^{g}(s,\cA)/\mbox{{\bf R}}.
\label{equivalence}\eeq
On the level of the representatives of the respective cohomology
classes, this isomorphism is established by \Gl{reconstr} for $q=n$.
\bsk

It should be stressed that the crucial property that makes (a) a
non-trivial descent go all the way down to a 0-form $\a\in{\cal A}$
and (b) the integration of the descent equations trivial, is
that $d\a$ is $s$-exact whenever $\a\in{\cal A}$ is BRST invariant.
This follows from \Gl{drepres1}, which reflects the presence
of the diffeomorphisms among the gauge symmetries, and
is of course not true for a generic gauge theory as the
example of electromagnetism shows. There the BRST transformation
of the gauge fields $A_\m$ and the ghost field $C$ are
$sA_\m=\6_\m C$ and $sC=0$. Hence, the field strengths $F_{\m\n}$ are
$s$-invariant but $dF_{\m\n}$ is not $s$-exact (for $n>2$).
As a consequence, in electromagnetism a non-trivial descent does not
necessarily terminate with a 0-form. A counterexample is
the abelian Chern-Simons 3-form $\a^{0,3}=AdA$ where $A=dx^\m A_\m$:
it is indeed non-trivial but the descent equations arising from
it terminate with a 2-form $\a^{1,2}=CdA$ due to $s\a^{0,3}+d\a^{1,2}=0$,
$s\a^{1,2}=0$. Furthermore, in contrast to
gravitational theories, there are non-trivial volume forms which
have a trivial descent, as e.g. $d^nx CF_{\m\n}$. Of course,
this does not exclude the existence of an operator $b$ which
satisfies \Gl{drepres1} on {\it particular}
solutions of the descent equations \cite{sorella}.

Nevertheless the descent equations that one encounters in
electromagnetism or Yang--Mills theory show up in gravitational
theories too, within a decomposition of the last equation $sa^{k+q,0}=0$ of
the gravitational descent equations into parts with definite degree
in the ghosts $\xi^\m$. In this decomposition the
$\xi^\m$ play a role similar to the differentials in the
descent equations of non-gravitational theories, such as
electromagnetism or Yang--Mills theory (indeed we will use a similar
decomposition in section \ref{scoh} to compute $H^*(s,\cA_2)$).
One may also take the point of view that it is the sum $dx^\m+\xi^\m$
which in gravitational theories takes the part of the differentials
in non-gravitational theories \cite{grav,Brandt,schweda,Baulieu}.

\mysection{BRST cohomology in $\cA_2$}\label{scoh}

\subsection{Notation}\label{notation}

In this subsection we prepare the computation and presentation
of $H^{*}(s,{\cal A}_2)$ by defining some functions which
we will encounter later.

First we introduce polynomials $\Th_\LAB$ in the ghosts
representing the so-called primitive
elements of the Lie algebra cohomology of $\cG$.
They correspond bijectively to the independent Casimir operators
of $\cG$ and therefore
their number equals the rank of $\cG$,
\[ \LAB=1,\ldots,rank(\cG).\]
One can construct the $\Th_\LAB$ explicitly by means
of suitable matrix representations $\{T^\LAB_I\}$ of $\cG$
\cite{orai,tal,lie},
\beq \Th_\LAB(\7C)=(-)^{m+1}\0 {m!(m-1)!}{(2m-1)!}
         \, tr(\7\cC_\LAB^{2m-1}),\qd m=m_\LAB,
         \qd \7\cC_\LAB=\7C^I T^\LAB_I
\label{r7}\eeq
where $m_\LAB$ denotes the order of the Casimir operator
corresponding to $\Th_\LAB$. Using \Gl{b9a} one can
extend \Gl{r7} to a Chern-Simons polynomial in the
ghosts $\7C^I$ and the curvature-quantities \Gl{b9b} \cite{lie},

\beq q_\LAB=
\sum_{k=0}^{m-1}(-)^k\0 {m!(m-1)!}{(m+k)!(m-k-1)!}\,
         Str(\7\cC\7\cB^k\7\cF^{m-k-1}),\qd \7\cB=\7\cC^2
\label{r5}\eeq
where $Str$ denotes the symmetrized trace of matrices
(cf. appendix \ref{ap1}) and on the r.h.s. it is understood that
\[ \7\cC=\7\cC_\LAB,\qd \7\cF=\7\cF_\LAB=\7F^IT^\LAB_I,\qd m=m_\LAB.\]
\Gl{r5} contains \Gl{r7} and
has been constructed such that its BRST transformation
does not depend on the $\7C^I$ at all,
\beq s\, q_\LAB=tr(\7\cF_\LAB^{m_\LAB})=:f_\LAB\ .
\label{sq}\eeq
Note that $f_\LAB$ vanishes for $m_\LAB> n/2$ since the $\7F^I$
are 2-``forms" in the $\7\xi^a$. The
(non-vanishing) $f_\LAB$ provide a basis
for the $\de_I$-invariant polynomials constructable of the $\7F^I$
(the $f_\LAB$ are not a basis in the vector space sense),
\beq \de_I P(\7F)=0\qd\LRA\qd P(\7F)=\cP(f).\label{invariants}\eeq
The $\Th_\LAB$ and $f_\LAB$ of the Lorentz group are explicitly
given in appendix \ref{ap2}. The abelian $\de_I\in\cG$
provide the $\Th_\LAB$ and $f_\LAB$ with
$m_\LAB=1$, given just by the abelian ghosts $\7C^I$ and
the abelian 2-``forms" $\7F^I$ respectively.

Next we introduce a basis $\{X^*_\De:\De=1,2,\ldots\}$ for
the vector space of inequivalent solutions of
\beq \de\, X^*_\De(\cT,\cT^*)= D_a J^a_\De(\cT),\qd
\de_I X^*_\De(\cT,\cT^*)=0,\qd
\agh{X^*_\De(\cT,\cT^*)}=1,\label{r9}\eeq
where two $X^*$'s are called equivalent if they differ
by functions of the form
$\de M(\cT,\cT^*)+D_a N^a(\cT,\cT^*)$ since the latter solve
\Gl{r9} trivially. Due to $[\de,\de_I]=0$ we can assume
with no loss of generality the $J^a_\De(\cT)$ to be
$\de_i$-invariant local functions transforming under $\cG_L$
according to the vector representation,
\beq \de_iJ^a_\De(\cT)=0,
\qd l_{ab}J_{c\De}(\cT)=
\et_{bc}J_{a\De}(\cT)-\et_{ac}J_{b\De}(\cT),\label{r9a}\eeq
since all pieces of the $J^a_\De$ which
do not satisfy \Gl{r9a} must result in contributions to
$D_aJ^a_\De$ which cancel separately.
We now define functions
$G^*_\De$ in terms of the solutions of \Gl{r9}
according to
\beq G^*_\De  = \lb X^*_\De -(-)^{\ep_\De}J^a_\De\0 \6{\6\7\xi^a}\rb\cE
\label{r11}\eeq
where $\ep_\De$ denotes the grading of $J^a_\De$
\footnote{With no loss of generality we assume
$X^*_\De$ and $J^a_\De$ to have definite (but opposite) grading.}
and $\cE$ plays the role of the volume element in $\cA_2$,
\beq\cE=\7\xi^0\ldots\7\xi^{n-1}=
\s0 {(-)^{n+1}}{n!}\ep_{a_1\ldots
a_n}\7\xi^{a_1}\ldots\7\xi^{a_n}.\label{vol}\eeq
$G^*_\De$ is BRST invariant because of \Gl{r9} and \Gl{r9a},
\beq s\, G^*_\De=0.\label{sG}\eeq
We anticipate that the $J_\De^a$ provide the ``covariantizable"
non-trivial local Noether currents, cf. section \ref{discussion}.
Therefore, if one knows the latter for the particular
model under consideration (i.e. for the particular choice of
the classical action), one can construct all
$G^*_\De$ explicitly (and vice versa).

Finally we introduce a label $\La$ for
those abelian factors of the gauge group
under which all
the matter fields $y^\Ly$ are uncharged,
\beq \{\de_\La\}=\{\de_I:\ [\de_I,\de_J]=0\ \forall J\ \wedge
\ \de_I y^\Ly=0\ \forall \Ly\}.\eeq
The corresponding gauge fields will be called
``free" henceforth although they generally couple to other fields
(in particular to the metric). The BRST transformation of the
antifields $C^*_{\La}$ following from \Gl{xx7} is given by
\beq s\, C^*_{\La}=\6_\mu(\xi^\mu C^*_{\La}-A^{*\mu}_{\La}).
\label{sc*}\eeq
This shows that the volume forms $d^nx C^*_{\La}$ are
solutions $a^{-2,n}$ of \Gl{i1}. In fact they are even non-trivial
according to \cite{bbh1}, section 13. Hence, they correspond
via descent equations (cf. section \ref{3}) to
BRST invariant 0-forms with ghost number $(n-2)$ which we denote
by $q^*_{\La}$,
\beq s\, q^*_{\La}=0. \label{sq*}\eeq
Contrary to \Gl{sc*}, the
explicit form of the $q^*_{\La}$ is not universal but
depends on the particular choice of
the classical action. However, from \Gl{sc*}
we can conclude that the $q^*_{\La}$
can be always chosen to be in ${\cal A}_2$ and
of the form
\beq q^*_{\La}=\7C^*_{\La}\, \cE
-\7A_{\La}^{*a}\0 {\6\cE}{\6\7\xi^a}+\ldots
\in {\cal A}_2 \label{r13}\eeq
where the non-written terms have degrees $\leq n-2$
in the $\7\xi$. For instance, \Gl{xx1} yields
\beq  q^*_{\La}
\simple\lb\7C^*_{\La}-\7A_{\La}^{*a}\0 \6{\6\7\xi^a}+\0 12\, F^{ab}_{\La}
\0 \6{\6\7\xi^a}\0 \6{\6\7\xi^b}\rb\, \cE.    \label{r13simple}\eeq

\subsection{Computation of $H^{*}(s,{\cal A}_2)$}\label{computation}

To compute $H^*(s,{\cal A}_2)$ we have to solve
\beq s\, \a=0,\qd \a\in\cA_2\label{m1}\eeq
modulo trivial solutions $s\be$ ($\be\in\cA_2$).
Two solutions are called equivalent if they differ by a trivial
solution. To solve \Gl{m1} we decompose it
into parts of definite degree
in the ghosts $\7\xi^a$ and call this degree {\it $\7\xi$-degree}.
The part of $\7\xi$-degree $p$ contained
in $\a$ is denoted by $\a_p$,
\beq \a=\sum_{p=\ell}^n\a_p,\qd N_{\7\xi}\a_p=p\a_p
\label{cc1}\eeq
where $\a_{\ell}\neq 0$ is understood and
$N_{\7\xi}$ denotes the counting operator for the $\7\xi$,
\[ N_{\7\xi}=\7\xi^a\0\6{\6\7\xi^a}.\]
The smallest $\7\xi$-degree
$\ell$ occurring in this decomposition is called the
{\it level} of $\a$. It provides a
simple but useful prescription for the choice
of representatives of the cohomology classes of
$H^*(s,{\cal A}_2)$:
out of two equivalent non-trivial
solutions of \Gl{m1} with different levels we choose always
the one with higher level.

The BRST operator decomposes into parts whose $\7\xi$-degrees
range from 0 to 2,
\beq s=s_0+s_1+s_2,\qd [N_{\7\xi},s_p]=ps_p.\label{m2}\eeq
The part $s_0$ is given by
\beq s_0=\de+\g_\cG,\label{s0}\eeq
where $\de$ is the Koszul--Tate differential whose action
in $\cA_2$ is given by \Gl{b8}, and $\g_\cG$ is the
longitudinal exterior derivative
along the gauge orbits of $G_L\times G_{YM}$
(``BRST operator of $G_L\times G_{YM}$"). The latter acts in
$\cA_2$ according to
\beq
\g_\cG=-\s0 12\7C^J\7C^K\, \f KJI\0 {\6}{\6\7C^I}
+\7C^I\de_I
\label{6}\eeq
where
\beq \de_i\7\xi^a=0, \qd
l_{ab}\7\xi_c=\et_{bc}\7\xi_a-\et_{ac}\7\xi_b,
\qd \de_I\7C^J=-\f IKJ\7C^K,
\label{m3}\eeq
i.e. explicitly one has
$\g_\cG \7C^I=\s0 12 \7C^J\7C^K\f KJI$ and
$\g_\cG Y=\7C^I\de_I Y$ for $Y\in\{\7\xi^a,\cT^r,\cT^*_\oor\}$.
The part $s_1$ acts on the generators of $\cA_2$ according to
\beq
s_1\,\cT^r=\7\xi^a D_a\cT^r,\qd
s_1\,\cT^*_\oor=\7\xi^a D_a\cT^*_\oor,\qd
s_1\,\7\xi^a=s_1\,\7C^I=0,
\label{s1}\eeq
i.e. $s_1$ is an exterior covariant derivative
in $\cA_2$ whose ``differentials" are the $\7\xi^a$ and
which vanishes on the $\7C^I$.
The part $s_2$ is non-vanishing only on the ghosts $\7C^I$ and
transforms the latter into the curvature-quantities \Gl{b9b}, i.e.
it is in $\cA_2$ given by the operator
\beq  s_2=\7F^I\0 {\6}{\6\7C^I}\ .\label{s2}\eeq
We note that $\de,\g_\cG,s_1$ and $s_2$ have trivial anti-commutation
relations apart from
\beq \g_\cG s_2+s_2\g_\cG+s_1^2=0.\label{11}\eeq

We obtain the following
decomposition of \Gl{m1} relative to the $\7\xi$-degree:
\bea 0&= & s_0\a_{\ell}\, ,\label{12a}\\
0&= &      s_0\a_{\ell+1}+s_1\a_{\ell}\, ,\label{12b}\\
0&= &      s_0\a_{p+2}+s_1\a_{p+1}+s_2\a_p\qd
                 \mbox{for}\qd\ell\leq p\leq n-2.\label{12c}\eea
Here we used that the $\7\xi$-degree cannot exceed $n$
since the $\7\xi$'s anticommute.

We shall now analyse \Gl{m1} starting from \Gl{12a}. To that
end we need the cohomology of $s_0$ in $\cA_2$:
\begin{lemma} The cohomology classes of $H^*(s_0,\cA_2)$ are
represented by
linearly independent polynomials in
the ghost-polynomials \Gl{r7} whose coefficient functions
are $\de_I$-invariant
polynomials in the $\7\xi^a$ and $\cT^r$,
\bea \lefteqn{s_0\a=0,\qd \a\in\cA_2
\qd\LRA \qd
\a=\a^i(\7\xi,\cT)\, \om_i(\7C)+s_0\be,}\nn\\
& & \de_I\a^i(\7\xi,\cT)=0,\qd
\om_i(\7C)=P_i(\Th(\7C)),\qd \be\in\cA_2.\label{m4}\eea
\label{LEM1}\end{lemma}

\proof{We decompose $\a$
into parts $\a^k$ with definite degree $k$
in the $\7C^I$ (``$\7C$-degree"), $\a=\sum_{k=0}^{\kappa}\a^k$.
Note that $\kappa$, the highest occurring $\7C$-degree, cannot
exceed $dim(\cG)$ as the $\7C^I$ anticommute.
Since the parts $\de$ and $\g_\cG$ of $s_0$
have $\7C$-degree 0 and 1
respectively, the condition $s_0\a=0$ implies in particular
$\g_\cG\,\a^{\kappa}=0$. Using
the Lie algebra cohomology of $\cG$
(cf. appendix \ref{ap2} for details), we conclude
$\a^\kappa=\5\a^{i_\kappa} \om_{i_\kappa}
+\g_\cG\5\be^{\kappa-1}$, $\de_I\5\a^{i_\kappa}=0$ where
$\5\a^{i_\kappa}=\5\a^{i_\kappa}(\7\xi,\cT,\cT^*)$,
and the $\om_{i_\kappa}=P_{i_\kappa}(\Th(\7C))$ are linearly
independent and
have $\7C$-degree $\kappa$. We now insert this result for $\a^\kappa$
in $\g_\cG\a^\kappa+\de\a^{\kappa+1}=0$, which is also implied by
$s_0\a=0$, and obtain
$(\de\5\a^{i_\kappa}) \om_{i_\kappa}
+\g_\cG(\a^{\kappa-1}-\de \5\be^{\kappa-1})=0$. Using the
Lie algebra cohomology of $\cG$ again, we conclude
$\de\5\a^{i_\kappa}=0$. The acyclicity of $\de$ in
positive antighost number (cf. \cite{henfisch} and appendices
\ref{ap4} and \ref{ap3} for details) implies $\5\a^{i_\kappa}=
\a^{i_\kappa}+ \de \be^{i_\kappa}$ where
$\a^{i_\kappa}=\a^{i_\kappa}(\7\xi,\cT)$ and
$\be^{i_\kappa}= \be^{i_\kappa}(\7\xi,\cT,\cT^*)$ are
$\de_I$-invariant polynomials. Now we consider
$\7\a:=\a-\a^{i_\kappa}\om_{i_\kappa}-s_0(
\5\be^{\kappa-1}+\be^{i_\kappa}\om_{i_\kappa})$. $\7\a$
is $s_0$-invariant due to $s_0\a=s_0(\a^{i_\kappa}\om_{i_\kappa})
=s_0^2=0$ and decomposes
into parts with $\7C$-degrees $<\kappa$. $\7\a$ is
now analysed as $\a$ and the arguments are
repeated until the highest occurring
$\7C$-degree has dropped to zero and one has proved the lemma.}

Now, we can always remove any piece of the form $s_0\be_{\ell}$
from $\a_\ell$ by subtracting the trivial solution
$s\be_\ell$ from $\a$ without lowering the level of $\a$.
Therefore we conclude from \Gl{12a} by means of lemma
\ref{LEM1} that we can assume with
no loss of generality
and in accordance with our prescription
for selecting representatives in $H^*(s,\cA_2)$ that
\beq \a_\ell=
\a^i_\ell(\7\xi,\cT)\, \om_i(\7C),\qd
\de_I\a^i_\ell(\7\xi,\cT)=0,\qd
\om_i(\7C)=P_i(\Th(\7C)).
\label{head}\eeq
Inserting \Gl{head} in \Gl{12b}, we obtain
$(s_1\a^i_\ell) \om_i+(\de+\g_\cG)\a_{\ell+1}=0$.
Now, $(s_1\a^i_\ell) \om_i$ does not contain any non-vanishing
$\g_\cG$-exact piece according to the
Lie algebra cohomology of $\cG$, cf. \Gl{liecohom2}
(note that $\de_I\a^i_\ell=0$ implies $\de_I(s_1\a^i_\ell)=0$).
Therefore we conclude that $\a_{\ell+1}$ contains a piece
$\a^i_{\ell+1} \om_i$ such that
\beq s_1\a^i_\ell+\de\a^i_{\ell+1}=0
\qd\LRA\qd s_1\a^i_\ell\approx 0.\label{head0a}\eeq
Furthermore we can assume
\beq \a^i_\ell\neq s_1\be^i_{\ell-1}+\de\be^i_\ell\qd\LRA\qd
\a^i_\ell\not\approx s_1\be^i_{\ell-1}
\label{head0b}\eeq
with no loss of generality since otherwise we can remove
the piece containing $\a^i_\ell$ from $\a_\ell$ by subtracting from $\a$
the trivial piece $s[(\be^i_{\ell-1}+\be^i_\ell)\om_i]$
(no sum over $i$ here). Note that this does not lower the level
of $\a$ since $\be^i_{\ell-1}$ and $\be^i_\ell$ have antighost number
0 and 1 respectively and are $\de_I$-invariant
with no loss of generality due to $\agh{\a_\ell^i}=\de_I\a_\ell^i=0$
and $[\de_I,s_1]=[\de_I,\de]=0$.

The solution of \Gl{head0a}, \Gl{head0b} is one of the crucial
steps within the computation. It is interesting in its own right since
it provides the weak equivariant cohomology of $d$ (cf. section
\ref{ass} for details). In fact it is an on-shell
version of the ``gravitational
covariant Poincar\'e lemma" of \cite{grav}. The latter provides
the (strong) equivariant cohomology of $d$ for Einstein--Yang--Mills
theories stating that the corresponding cohomology classes
are in form degrees $<n$ represented by invariant polynomials
in the curvature 2-forms $R^{ab}=\s0 12 dx^\m dx^\n {R_{\m\n}}^{ab}$
and $F^i=\s0 12 dx^\m dx^\n {F_{\m\n}}^i$.\footnote{Analogous results
have been derived already in \cite{Gilkey} for
pure gravity (Riemannian manifolds) and in \cite{com}
for Yang--Mills theory.} Recalling the results of \cite{bbh1} for the
characteristic cohomology of $d$ (in the space of all local forms),
it is not too surprising that in all
form degrees $<n-2$ the weak and strong equivariant cohomology
of $d$ agree whereas they differ in form degrees $\geq n-2$:
\begin{theorem}\label{CPL}
(Weak covariant Poincar\'e lemma)

(i) $\de_I$-invariant functions $\a(\7\xi,\cT)$ which are
weakly $s_1$-closed are weakly $s_1$-exact up to invariant
polynomials in the $\7F^I$,
functions of maximal $\7\xi$-degree and
the antifield independent
parts of the $G^*_\De$ and of those $q^*_{\La}$ which
do not depend on the $\7C^I$,
\bea & &s_1\a(\7\xi,\cT)\approx 0,\qd
\de_I\a(\7\xi,\cT)=0\nn\\
&\LRA & \a\approx
P(\7F)+\la^{\La^0}q^*_{\La^0}|_{\PH^*=0}+
\la^{\De}G^*_{\De}|_{\PH^*=0} +\cE L(\cT)+s_1\be
\label{CPLi}
\eea
where $\be=\be(\7\xi,\cT)$ and
\beann & & \de_IP(\7F)=\de_I L(\cT)=\de_I\be(\7\xi,\cT)=0,\\
& &\{q^*_{\La^0}\}=\{q^*_{\La}:\ \6 q^*_{\La}/\6\7C^I=0\}.
\eeann

(ii) No non-vanishing $\de_I$-invariant polynomial
$P(\7F)$ with ghost number
$<n-1$ is weakly $s_1$-exact,
\beq d_{I_1\ldots I_k}\7F^{I_1}\ldots \7F^{I_k}\approx
s_1\be,\qd 2k<n-1\qd\LRA\qd d_{I_1\ldots I_k}=0\label{CPLii}\eeq
where the $d_{I_1\ldots I_k}$ denote totally symmetric
constant tensors of $\cG$. \Gl{CPLii} remains valid for
$2k=n-1$ if all $q^*_\La$ can be assumed
to independent of the $\7C^I$, i.e. if $\{q^*_\La\}=\{q^*_{\La^0}\}$.
\end{theorem}

The proof of this important theorem is given in appendix \ref{B}.
Here we just note that, apart from a refined version of
the above-mentioned results derived in \cite{bbh1},
the part $s_2$ of the BRST operator plays an important role
within the proof. By restricting the proof to antifield
independent functions, it provides the strong $s_1$-cohomology, i.e.
an alternative (improved) proof of the (strong) covariant
Poincar\'e lemma which does not use the
split $\e \m{a}=\de_\m^a+h_\m^a$, contrary to the original proof
given in \cite{grav}.

We shall now complete the analysis assuming that
there are no ``free" abelian gauge fields and that
\Gl{CPLii} holds even for $2k=n$,
\beq 2k=n:\qd
d_{I_1\ldots I_k}\7F^{I_1}\ldots \7F^{I_k}\approx s_1\be
\qd\LRA\qd d_{I_1\ldots I_k}=0.\label{CPLiii}\eeq
That is, we assume that the characteristic classes
of degree $n$ remain non-trivial
in the equivariant cohomology of $d$ even on-shell.
In section \ref{volume} we comment this assumption and
illustrate that it is physically reasonable. We also
show there that it is equivalent to the absence
of Chern-Simons forms in (non-trivial)
Noether currents. We can now prove:
\begin{theorem}
If there are no ``free" abelian gauge fields and if
\Gl{CPLiii} is valid, then one has
\beq
s\a=0,\ \a\in\cA_2\qd\LRA\qd
\a=s\be+\cE L_\tau(\cT) \om^\tau(\7C)+P(q,f)+G^*_\De \om^\De(\7C)
\label{theo1}
\eeq
where the $L_\tau(\cT)$ are $\de_I$-invariant polynomials in
the $\cT^r$, $\om^\tau(\7C)$ and $\om^\De(\7C)$ are
polynomials in the ghost polynomials \Gl{r7}, and $P(q,f)$
are BRST invariant polynomials in the ``Chern--Simons polynomials"
\Gl{r5} and the ``Chern polynomials" \Gl{sq},
\beq \de_IL_\tau(\cT)=0,\qd \om^\tau(\7C)=P^\tau(\Th(\7C)),\qd
\om^\De(\7C)=P^\De(\Th(\7C)),\qd sP(q,f)=0.\label{theo1a}\eeq
Furthermore, under the same assumptions as above (absence of
``free" abelian gauge fields and validity of \Gl{CPLiii})
a polynomial $P(q,f)$ is $s$-exact if and only if it is
$s$-exact in the space of polynomials in the $q$'s and $f$'s,
\beq P(q,f)=s\be\qd\LRA\qd P(q,f)=sQ(q,f).\label{theo1ab}\eeq
\label{THEO}
\end{theorem}

\proof{Assume that $\a$ represents a
non-trivial cohomology class of $H^*(s,\cA_2)$ and
has level $\ell$. Then its part $\a_\ell$ has the form \Gl{head}
where the $\a_\ell^i$ are determined by theorem
\ref{CPL}.

Since by assumption there are no $q^*_\La$,
part (i) of theorem \ref{CPL} implies
in all cases $\ell\leq n-2$
that we can assume the $\a_\ell^i$ to be invariant
polynomials in the $\7F^I$ (which implies that $\ell$ is even),
\beq\ell\leq n-2:\qd \a_\ell= P^i_\ell(\7F)\, \om_i(\7C),\qd
\de_I P^i_\ell(\7F)=0,\qd \om_i(\7C)=P_i(\Th(\7C)).
\label{comp1}\eeq
We complete now all
$\Th_\LAB(\7C)$ occurring in \Gl{comp1} to the corresponding $q_\LAB$
introduced in \Gl{r5} and consider
\beq \be:=\a-P^i_\ell(\7F)\, P_i(q).\label{comp2}\eeq
Note that, by construction, $\be$ has higher level than
$\ell$,
\beq \be=\sum_{p\geq \ell(\be)}\be_p,\qd
N_{\7\xi}\be_p=p\be_p,\qd \ell(\be)>\ell\ .\label{przz6}\eeq
Since $\a$ is BRST invariant (by assumption),
the BRST transformation of $\be$ is given by
\beq s\be=-s\, [P^i_\ell(\7F)\, P_i(q)]
=:-\cP(q,f)\label{comp3}\eeq
where we used \Gl{sq} and \Gl{invariants}. Since $s$ contains no
part with negative $\7\xi$-degree, $\cP(q,f)$ has level
$p_0\geq \ell(\be)$,
\beq \cP(q,f)=\sum_{p\geq p_0}\cP_p,\qd N_{\7\xi}\cP_p
=p\cP_p,\qd p_0\geq \ell(\be).\label{przz2}\eeq
\Gl{comp3} implies
\bea -\cP_{p_0}&=&s_0\be_{p_0}+s_1\be_{p_0-1}+s_2\be_{p_0-2},
\label{przz7}\\
0&=&s_0\be_{p_0-1}+s_1\be_{p_0-2}+s_2\be_{p_0-3},
\label{przz8}\\
&\vdots&\nn\\
0&=&s_0\be_{\ell(\be)+1}+s_1\be_{\ell(\be)},\label{przz9}\\
0&=&s_0\be_{\ell(\be)}.\label{przz10}\eea
We distinguish
the cases $\ell(\be)\geq p_0-1$ and $\ell(\be)< p_0-1$.
Assume first that $\ell(\be)\geq p_0-1$. Then \Gl{przz7}--\Gl{przz10}
reduce to
\bea -\cP_{p_0}=
s_0\be_{p_0}+s_1\be_{p_0-1},\qd
0=s_0\be_{p_0-1}\, .\label{przzz16}\eea
Now, $\cP_{p_0}$ is of the form
\[ \cP_{p_0}=P^i_{p_0}(\7F)\om_i(\7C),\qd
\om_i(\7C)=P_i(\Th(\7C))\]
where the $P^i_{p_0}(\7F)$ are $\de_I$-invariant
polynomials in the $\7F^I$ with ghost number ${p_0}$.
Using lemma \ref{LEM1} and the Lie algebra cohomology
of $\cG$, we conclude from \Gl{przzz16}
that $\be_{p_0-1}$ and
$\be_{p_0}$ must
contain parts $\be^i_{p_0-1}(\7\xi,\cT)\om_i(\7C)$ and
$\be^i_{p_0}(\7\xi,\cT,\cT^*)\om_i(\7C)$
such that
\beq P^i_{p_0}(\7F)=s_1\, \be^i_{p_0-1}(\7\xi,\cT)+
\de\, \be^i_{p_0}(\7\xi,\cT,\cT^*)
\ .\label{przz17}\eeq
Since this is equivalent to $P^i_{p_0}(\7F)\approx
s_1\, \be^i_{p_0-1}$,
part (ii) of theorem \ref{CPL} and assumption
\Gl{CPLiii} imply $P^i_{p_0}(\7F)=0$ in \Gl{przz17}
(note: this is the place where we use \Gl{CPLiii}). We
conclude $\cP(q,f)=0$ and thus that $\be$ is
BRST invariant, cf. \Gl{comp3}. \Gl{comp2} implies now
that $\a$ agrees with $P^i_\ell(\7F)\, P_i(q)$
up to a separately BRST invariant function $\be$ which
has higher level than $\a$.

In the case $\ell(\be)< p_0-1$ we conclude from \Gl{przz9} and
\Gl{przz10}, repeating the arguments that have lead to
\Gl{head}, \Gl{head0a} and \Gl{head0b} and
using part (i) of theorem \ref{CPL}, that we can assume
with no loss of generality
(by subtracting $s$-exact pieces from $\be$ if necessary)
\beq \ell(\be)< p_0-1:\qd \be_{\ell(\be)}=
P_{\ell(\be)}^i(\7F)\om_i(\7C)=:
{\cal Q}(\Th(\7C),f)\label{przz11}\eeq
since $\ell(\be)<p_0-1$ implies in particular $\ell(\be)<n-1$.
We now complete the $\Th_\LAB(\7C)$ occurring in ${\cal Q}$ to
the corresponding $q_\LAB$ and consider the
function $\be':=\be-{\cal Q}(q,f)$ whose
level $\ell(\be')$ exceeds $\ell(\be)$ by construction. Since
$s{\cal Q}(q,f)=\cP'(q,f)$ is again a polynomial in the
$q_\LAB$ and $f_\LAB$ due to \Gl{sq}, we get
\[s\, \be=\cP'(q,f)+s\, \be',\qd \ell(\be')>\ell(\be).\]
\Gl{comp3} can now be rewritten in the form
\beq s\, \be'=-\cP(q,f)-\cP'(q,f).
\label{przz12}\eeq
We can analyse \Gl{przz12} as \Gl{comp3} before and repeat the arguments
until we arrive at
\beq s\, \7\be=-\7\cP(q,f),\qd \7\be=\be-{\cal Q}(q,f)-{\cal Q}'(q,f)-\ldots,
\qd \ell(\7\be)\geq \7p_0-1
\label{przz13}\eeq
where $\ell(\7\be)$ and $\7p_0$ are the levels of $\7\be$
and $\7\cP(q,f)=\cP(q,f)+\cP'(q,f)+\ldots$ respectively.
This traces the case $\ell(\be)< p_0-1$
made in the distinction after \Gl{przz10} back
to the case $\ell(\be)\geq p_0-1$ and we conclude
$\a=P^i_\ell(\7F)\, P_i(q)+{\cal Q}(q,f)+{\cal Q}'(q,f)+\ldots$
up to a separately BRST invariant function $\7\be$ whose
level exceeds $\ell$.

We conclude that we can assume
$\a=P(q,f)$ in the cases $\ell \leq n-2$, up to
separately BRST invariant functions whose level exceed $\ell$. The
latter represent different cohomology classes in accordance with
our rule for selecting representatives of the cohomology
classes of $H^*(s,\cA_2)$ (cf. beginning of this subsection).
The arguments used within the discussion of \Gl{comp3} show
also that
$\cP(q,f)=s\be$ implies $\be=\7\be+{\cal Q}(q,f)+{\cal Q}'(q,f)+\ldots$
where $\7\be$ is
separately $s$-invariant. This proves \Gl{theo1ab}.

Consider finally the cases $\ell=n-1$ and $\ell= n$.
Part (i) of theorem \ref{CPL} implies that
we can assume
\bea \ell=n-1: && \a_{n-1}=
P^i_{n-1}(\7F)\om_i(\7C)+G^*_\De |_{\PH^*=0} \om^\De(\7C)
\label{comp5}\\
\ell=n: && \a_{n}=\cE\, L_\tau(\cT)\, \om^\tau(\7C)
\label{comp6}\eea
where the $\om(\7C)$ are polynomials in the ghost polynomials
\Gl{r7} and
 $P^i_{n-1}(\7F)$ and $L_\tau(\cT)$ are
$\de_I$-invariant polynomials (invariant polynomials $P_{n}(\7F)$
count among the $L_\tau(\cT)$).
Now, $s\Th_\LAB(\7C)=s_2\Th_\LAB(\7C)$ has $\7\xi$-degree 2.
Since $P^i_{n-1}(\7F)$, $G^*_\De$ and $\cE L_\tau(\cT)$ are
BRST invariant and have levels $\geq n-1$, we conclude that
\Gl{comp6} is already BRST invariant and that \Gl{comp5} can
be completed to the BRST invariant function
$\7\a=P^i_{n-1}(\7F)\om_i(\7C)+G^*_\De\om^\De(\7C)$. This proves the
theorem since it implies in the case $\ell=n-1$ that
$\a$ agrees with $\7\a$ up to a BRST invariant function with
level $n$ which is of the form \Gl{comp6} modulo an $s$-exact piece.}

Thanks to theorem \ref{THEO}
one can now complete the analysis by
a computation of the BRST cohomology in the space
of the polynomials $P(q,f)$ only.
This has been done in \cite{tal,lie} and we refer to these
references for the result. We note that
functions $\cE L(\cT) P(\Th(\7C))$ are trivial if
$L(\cT)\approx D_a M^a(\cT)$ holds for some $M^a(\cT)$.

Let us finally indicate the modification of the result for
$H^*(s,\cA_2)$ in presence of
``free" abelian gauge fields. The latter give rise to additional
solutions of \Gl{m1} with level $\ell=n-2$ due to theorem \ref{CPL}
and make it necessary to consider this
case separately (all other values of $\ell$ can be
treated as within the proof of theorem \ref{THEO}). The representatives
of the resulting additional
cohomology classes contain the $q^*_\La$ and have to be evaluated
for each classical action separately since the latter determines
the explicit form of the $q^*_\La$. We just note that for the
action \Gl{xx1} the additional cohomology classes are represented
by functions
\beq \a_{add}\simple \sum_\La
q^*_{\La}\0 {\6P(q)}{\6\7C^\La}  \label{additional}\eeq
where the polynomials $P(q)$ depend only on the $q_\LAB$ of the semi-simple
part of $\cG$ and on the ``free" abelian ghosts $\7C^\La$, but not
on other abelian ghosts.

\mysection{Equivariant characteristic cohomology}\label{ass}

We will now show that theorem \ref{CPL} provides indeed
the equivariant characteristic cohomology.
We define the latter as the weak cohomology of $d$
in the space of local $p$-forms
which (a) truly transform as $p$-forms under spacetime
diffeomorphisms, (b) are invariant
under local Lorentz and Yang--Mills gauge
transformations, and (c) do not depend on the ghosts, the
antifields and (explicitly) on the $x^\m$. Denoting such $p$-forms by
\beq a^{p}_{inv}= \s0 1{p!}\, dx^{\m_1}\ldots dx^{\m_p}
\a_{\m_1\ldots\m_p}\ ,\label{ch0}\eeq
the requirements (a)--(c) impose that
the BRST transformation of the local coefficient
functions $\a_{\m_1\ldots\m_p}$ coincides with
their standard Lie derivative
along the diffeomorphism ghost,
\bea s\a_{\m_1\ldots\m_p}&=&\cL_\xi \a_{\m_1\ldots\m_p}\nn\\
&=& \xi^\n\6_\n\a_{\m_1\ldots\m_p}
+\sum_{i=1}^p(\6_{\m_i}\xi^\n)\a_{\m_1\ldots\m_{i-1}\n\m_{i+1}\ldots\m_p}\ .
\label{ch1}\eea
The absence of derivatives of the ghosts in
\Gl{ch1} apart from the first order derivatives of the $\xi^\m$
implies already that $\a_{\m_1\ldots\m_p}$
can be written entirely in terms of the tensor fields
\Gl{b1} and
the undifferentiated $\e \m{a}$. This is most easily proved using
the basis of generators of $\cA$ introduced in section \ref{basis}.
Indeed, \Gl{ch1} implies that no generator \Gl{b4} occurs
in $sa^p_{inv}$ except for those given by $s\e \m{a}$ and
thus that $a^p_{inv}$ itself does not depend on the generators
\Gl{b3} apart from the undifferentiated $\e \m{a}$.
The absence of the undifferentiated ghosts
$C^I$ in \Gl{ch1} then
imposes in addition that $\a_{\m_1\ldots\m_p}$ is $\de_I$-invariant
and we conclude
\bea  \a_{\m_1\ldots\m_p}=\e {\m_1}{a_1}\ldots\e {\m_p}{a_p}
\a_{a_1\ldots a_p}(\cT) ,\qd
\de_I\a_{\m_1\ldots\m_p}=0\ .\label{ch2}\eea
Hence, \Gl{ch1} implies \Gl{ch2}. Conversely,
\Gl{ch2} implies \Gl{ch1} and is therefore equivalent to the latter.
Note that \Gl{ch2} allows us to express $a^{p}_{inv}$ in the form
\beq a^{p}_{inv}= \s0 1{p!}\, e^{a_1}\ldots e^{a_p}
\a_{a_1\ldots a_p}(\cT)\label{ch0a}\eeq
where the $e^a$ are the vielbein 1-forms
\beq e^a=dx^\m \e \m{a}.\label{ch0b}\eeq
Due to \Gl{0} these 1-forms satisfy
\beq D e^a=0\label{ch0c}\eeq
where $D$ denotes the exterior covariant derivative
\beq D=e^b D_b=dx^\m (\6_\m-\A \m{I}\de_I).\label{ch0d}\eeq
Note that we have defined $D$ without a piece containing the
Christoffel connection ${\Gamma_{\m\n}}^\rho$. It is
not necessary to include such a piece here since it
drops out when $D$ acts on any $a^p_{inv}$. In fact
even the piece $dx^\m \A \m{I}\de_I$ drops out due to
\Gl{ch2}. Thus $D$ actually agrees
with $d$ on the forms $a^{p}_{inv}$,
\beq Da^{p}_{inv}=da^{p}_{inv}.\label{ch0e}\eeq
It is now evident that the
equivariant characteristic cohomology is indeed
described by theorem \ref{CPL} since
$D$ turns into $s_1$ on functions of the $\cT^r$ and $e^a$
when one substitutes $e^a\rightarrow \7\xi^a$
and vice versa (recall that $s_1\7\xi^a=0$, which
is analogous to \Gl{ch0c}). Note that
this shows again that the diffeomorphism ghosts
play a role similar to the differentials, cf. remarks
at the end of section \ref{3}. We conclude from
theorem \ref{CPL} that
the cohomology classes of the
equivariant characteristic cohomology
are in form degrees $<n-2$
represented by characteristic
classes\footnote{By an abuse of
terminology we call the invariant polynomials
in the curvature 2-forms characteristic classes.},
i.e. by independent invariant polynomials in the
curvature 2-forms
\[ F^i=\s0 12 dx^\m dx^\n {F_{\m\n}}^i,\qd
F^{ab}=\s0 12 dx^\m dx^\n {R_{\m\n}}^{ab}.\]
This implies in particular
that the equivariant characteristic cohomology is
zero in all odd form degrees $<n-2$,
\beq\ba{llcl} 2k<n-3:& da^{2k+1}_{inv}\approx 0& \LRA &
a^{2k+1}_{inv}\approx db^{2k}_{inv}\ ,\\
2k<n-2:& da^{2k}_{inv}\approx 0 & \LRA &
a^{2k}_{inv}\approx d_{I_1\ldots I_k}F^{I_1}
\ldots F^{I_k}+db^{2k-1}_{inv}\  .
\ea\label{ch8}\eeq
Moreover, part (ii) of theorem \ref{CPL} implies
that no characteristic class
with form degree $<n-1$ is trivial in the equivariant
characteristic cohomology,
\beq d_{I_1\ldots I_k}F^{I_1} \ldots F^{I_k}\approx
db^{2k-1}_{inv},\qd 2k<n-1\qd\then\qd d_{I_1\ldots I_k}=0.
\label{ch9}\eeq
We shall comment this result at the end of this section.

We note that \Gl{ch8} and \Gl{ch9} remain valid for strong equalities,
cf. \cite{grav,Gilkey} and the remarks
preceding and following theorem 7.1.
That is, the strong and the weak equivariant cohomology of $d$
agree in all form degrees $<n-2$. This is not true
in form degrees $(n-2)$ and $(n-1)$ since in these cases
the equivariant characteristic cohomology contains in
general cohomology classes which cannot be represented
by invariant polynomials in the
curvature 2-forms (the
strong equivariant cohomology of $d$ is in form degrees
$(n-2)$ and $(n-1)$ still represented only
by characteristic classes).
In degree $(n-2)$ there is a difference only if ``free" abelian
gauge fields are present. They give rise to additional
cohomology classes corresponding to the
$q^*_{\La^0}$ occurring in \Gl{CPLi}.
The explicit form of these $(n-2)$-forms depends
on the particular classical action. For the standard Lagrangian
\Gl{xx1} one has $\{q^*_{\La^0}\}=\{q^*_{\La}\}$ and obtains simply
\beq a^{n-2}_{\La}\simple\s0 {1}{2(n-2)!}(-)^{n+1} dx^{\m_1}\ldots
dx^{\m_{n-2}}\ep_{\m_1\ldots\m_n}F^{\m_{n-1}\m_n}_{\La}.
\label{ch-1}\eeq
The equivariant characteristic cohomology
in form degree $(n-1)$ is represented by
characteristic classes of this degree (if any) and
by forms corresponding to the
$G^*_\De$ according to \Gl{CPLi}. The latter are given by
\beq
a^{n-1}_{inv,\De}= \s0 1{(n-1)!}\, dx^{\m_1}\ldots dx^{\m_{n-1}}
\ep_{\m_1\ldots \m_n}\E a{\m_n}J^a_\De(\cT).
\label{ch10}\eeq
These $(n-1)$-forms are the covariantized Noether currents,
cf. section \ref{discussion}.

Let us finally comment the above results.
As we have seen, the strong and weak equivariant
cohomology of $d$ can be different only in form degrees $\geq n-2$.
This has a deep reason given by the irreducibility
of the gauge symmetry. In fact
it traces back to an analogous result holding for
the characteristic cohomology itself. Indeed we have
shown in \cite{bbh1} that the characteristic cohomology
vanishes in all non-vanishing
form degrees $<n-2-r$ for any normal gauge
theory of reducibility order $r$ if the field-antifield
manifold is contractible. In our case this remains true
except for the occurrence of the topological conserved forms
$\a_m^{0,k_m}$, cf. section \ref{discussion}. These forms
reflect the non-trivial De Rham cohomology
of the vielbein manifold, cf. section \ref{APL},
but do not show up in the
equivariant characteristic cohomology since they do not
satisfy \Gl{ch1}.

In particular,
the irreducibility of the gauge symmetry underlies
the result that the characteristic classes of degree $<n-1$
remain non-trivial in the equivariant
cohomology of $d$ even on-shell.
This is not true, in general, for reducible gauge
theories. Indeed, a counterexample is
provided by the ``$BF$-theory" defined through the following
Lagrangian
\beq \cL=B_{n-2}F,\qd
B_{n-2}=\s0 1{(n-2)!}dx^{\m_1}\ldots dx^{\m_{n-2}}
        B_{\m_1\ldots{\m_{n-2}}},\qd
F=\s0 12 dx^\m dx^\n \6_\m A_\n\ .\label{BF}\eeq
Evidently, this defines for $n>3$ a reducible gauge theory  of
reducibility order $r=n-3$
and implies that the abelian characteristic class $F$ vanishes
on-shell. Hence, \Gl{ch9} does not hold in
general for reducible gauge theories.

Contrary to the characteristic classes of form degrees $<n-1$,
those of form degrees $(n-1)$ or $n$ can be trivial
in the equivariant characteristic cohomology even in irreducible
gauge theories.
Part (ii) of theorem \ref{CPL} shows that this can happen
in the case $(n-1)$ only in the academical context that
``free" abelian gauge fields are present. Indeed, the abelian
Chern--Simons theory given by \Gl{BF} for $n=3$
provides an example of this phenomenon. That
characteristic classes of form degree $n$ can be trivial in the
equivariant characteristic cohomology is
also illustrated by \Gl{BF}, for $n=2$. A similar example in
four dimensions is discussed at the end of section \ref{volume}.

\mysection{Result for $H^{*,n}(s|d)$}\label{volume}

According to section \ref{3}
it is now straightforward to construct
$H^{*,*}(s|d)$.
We only spell out $H^{*,n}(s|d)$ in some detail
since it contains the solutions of direct
physical interest. The discussion
of these solutions is relegated to section \ref{discussion}.
We will assume that \Gl{CPLiii} holds. At the end
of the section we will illustrate that \Gl{CPLiii}
is a physically reasonable assumption and
indicate the changes in the results if it
is not satisfied.

The solutions $a^{g,n}$ of \Gl{i1} arise
via the descent equations from the functions
\Gl{expansion2} with ghost number $(g+n)$ according to
section \ref{3}. One can distinguish three classes of them,
stemming solely from $\cA_2$, solely from $\cA_1$, or from
functions \Gl{expansion2} involving only non-constant
representatives of $H^*(s,\cA_i)$.

Those of the first class stem from the cohomology classes
of $H^*(s,\cA_2)$. If ``free" abelian gauge fields are absent and
assumption \Gl{CPLiii} holds, this gives
three different types of solutions
$a^{g,n}_i$ ($i=1,2,3$) of \Gl{i1} arising from
corresponding polynomials occurring
in \Gl{theo1} respectively,
\beq\ba{ccc}
\cE L_\tau(\cT) P^\tau(\Th(\7C))&\rightarrow & a^{g,n}_1 ,\\
P(q,f)&\rightarrow & a^{g,n}_2,\\
G^*_\De P^\De(\Th(\7C))&\rightarrow & a^{g,n}_3.\ea\label{types1-3}\eeq
The solutions $a^{g,n}_1$ and $a^{g,n}_2$ do not
involve antifields at all and are well-known in the
literature. They have been denoted by $\cA_{trace}$ and
$\cA_{chiral}$ respectively in \cite{grav} and are given there
explicitly in eq. (7.42). We do not spell them out again,
but just note that among them there
are the invariant actions, as well as the chiral anomalies.

The solutions
$a^{g,n}_3$ involve non-trivially the
antifields and were previously unknown.
According to section \ref{3} they are
given by the $n$-forms contained
in $G^*_\De P^\De(\Th(\7C))$
after replacing $\xi^\mu$ by $dx^\mu$.
When doing so, one must not forget that the ghosts $\7C^I$ involve
the $\xi^\mu$ due to \Gl{b2} which gives
\[  P(\Th(\7C))=(1+\xi^\m \A \m{I}\0 \6{\6 C^I}+\ldots)\,
P(\Th(C))\]
where the non-written terms are at least quadratic in the $\xi^\mu$.
Using \Gl{r11}, we get
\beq a^{g,n}_3= d^nx\, e \lb
X^*_\De+(-)^{\ep_\De}J^\m_\De
\A \m{I}\0 \6{\6C^I}\rb P^{\De,g+1}(\Th(C))
\label{n6}\eeq
with $J^\m_\De=\E a\m J^a_\De$. The
polynomials $P^{\De,g+1}(\Th(C))$ have ghost number
$(g+1)$ since $a^{g,n}_3$ is supposed to have ghost number $g$.
The factor $e$ in \Gl{n6}
originates from the substitution $\xi^\mu\rightarrow dx^\mu$ since the
latter results in $\cE\rightarrow d^nx\, e$. The different signs in
front of the $J$-terms in \Gl{n6} and \Gl{r11} stem from the grading
of the volume element.
Note that in the important cases $g=0,1$ the polynomials
$P^{\De,g+1}(\Th(C))$ have ghost number 1 and 2 respectively and
therefore involve only abelian ghosts because the $\Th$'s
of the semi-simple parts of $\cG$ have ghost numbers $\geq 3$.
Hence, there are no solutions
$a^{g,n}_3$ with $g=0,1$
if $G_{YM}$ is semi-simple  (contrary to other values of $g$).

If ``free" abelian gauge fields are present, then the first class
of solutions contains additional
solutions of a fourth type. They stem from
cohomology classes of $H^*(s,\cA_2)$ whose representatives
involve the $q^*_\La$
and therefore depend on antifields, cf.
end of subsection \ref{computation}.
Their explicit form depends on the particular
action. For the standard Lagrangian \Gl{xx1},
one obtains from \Gl{r13simple}
and \Gl{additional}
\beq a^{g,n}_4\simple  d^nx \sum_\La\lb C^*_{\La}+
\As {\La}\m\A \m{I}\0 \6{\6C^I}+
\0 e2\, F^{\m\n}_{\La}{\cal Z}_{\n\m}(C,A,F)\rb
\0 {\6P^{g+3}(\Th(C))}{\6C^\La}
\label{n7}\eeq
where the $P^{g+3}(\Th(C))$ have ghost number
$(g+3)$ and depend on the $\Th$'s of the semi-simple part of $\cG$
and on the ``free" abelian ghosts $C^\La$, but not on other abelian ghosts.
The ${\cal Z}_{\m\n}$ are
field dependent operators
\beq {\cal Z}_{\m\n}(C,A,F)=\A \m{J}\A \n{I}\0 \6{\6C^I}\0 \6{\6C^J}+
 \F \m\n{I}q_{I,\LAB}(C)\0 \6{\6\Th_{\LAB}(C)} \label{defz}\eeq
with
\beq  q_{I,\LAB}(C)=
(-)^{m}\0 {m!(m-1)!}{(2m-2)!}\, tr(T^{\LAB}_I\, \cC^{2m-3}),
\qd m=m_{\LAB},\qd \cC=C^IT^{\LAB}_I.\label{defq}\eeq
[$q_{I,\LAB}(C)$ vanishes for $m_{\LAB}=1$].
We note that \Gl{n7}
agrees in fact with equation (8.17) of \cite{bbh2} up to the factor
$e$ in front of the $F^{\m\n}_{\La}$-terms. The total antisymmetry
of the coefficients $f$ occurring in equation (8.17) of \cite{bbh2}
(in their $\a$-indices) is
provided in \Gl{n7} by the differentiation of $P^{g+3}$ with respect
to $C^\La$ (and by the odd grading of the abelian ghosts contributing
to $P^{g+3}$).
\bsk

The solutions of the second class arise from
$H^*(s,\cA_1)$ and are denoted by $a^{g,n}_{5}$. They
are simply the forms $\a^{k_m-n,n}_m$ corresponding
to $\a^{k_m,0}_m$ via the descent equations (cf. section \ref{2}),
\beq a^{g,n}_{5}=\sum_{\{m|k_m=n+g\}}\la_{m}\a^{k_m-n,n}_m\label{top1}\eeq
where $\la_m$ are constants.
Note that these solutions are present only for particular values of
$g$ and $n$ since $k_m$ takes only special values
obtained from the Lie algebra cohomology of $so(n)$, cf.
section \ref{APL} and appendix \ref{ap2}.

The solutions of the third class arise from
functions \Gl{expansion2}
involving only non-constant $\a^{k_m,0}_m$'s and $P_A$'s.
We denote them by $a^{g,n}_{6}$.
The non-constant $P_A$ have
ghost numbers $\geq n-2$ according to appendix \ref{B}
(lemma \ref{lem7}). Since the non-constant $\a^{k_m,0}_m$
have ghost number $k_m\geq 3$, we conclude
that solutions $a^{g,n}_{6}$ exist only for $g\geq 1$.
Of direct physical interest are only those with $g=1$ since
they are candidate anomalies.
They arise solely from the $\a^{k_m}_m$ with $k_m=3$
and the $P_A$ with ghost number $(n-2)$. The latter are
just the $q^*_\La$, cf. lemma \ref{lem7}. We get
\beq a^{1,n}_{6}=\sum_\Lb \sum_{\{m|k_m=3\}}\sum_{i=0}^3 \la_{m,\Lb}
\a^{k_m-i,i}_m a^{i-2,n-i}_\Lb\label{top2}\eeq
where $\la_{m,\Lb}$ are constants and
the $a^{i-2,n-i}_\Lb$ are the $(n-i)$-forms occurring in the descent
equations corresponding to the $q^*_\Lb$, i.e.
\[ a^{-2,n}_\Lb=d^nx C^*_\Lb\ ,\qd\ldots\qd,\qd a^{n-2,0}_\Lb=q^*_\Lb\ .\]

Note that the solutions $a^{g,n}_4$ and
$a^{1,n}_{6}$ are present only in the academical case
that there are ``free" abelian gauge fields. In particular there
are no such solutions if $G_{YM}$ is semi-simple.

Let us finally indicate modifications in the results
if \Gl{CPLiii} is not valid. To illustrate them
we consider a four dimensional theory whose
Lagrangian has the standard form \Gl{xx1} up to the
following extra piece:
\beq n=4:\qd \cL^\ph/e=\s0 12\, g^{\m\n}(\6_\m \ph)\,(\6_\n \ph)
-\s0 14\ph\, \ep^{\m\n\rho\si}{F_{\m\n}}^I {F_{\rho\si}}^J d_{IJ}.
\label{cp4}\eeq
Here $\ph$ is a real uncharged scalar field which
for simplicity is assumed not to occur
in the original Lagrangian \Gl{xx1}. We note that the sum of
\Gl{cp4} and \Gl{xx1} still defines a normal theory in the
terminology of \cite{bbh1}. One easily verifies that
\Gl{cp4} implies
\beq F^I F^Jd_{IJ}\approx -db^3_{inv},\qd
b^3_{inv}=\s0 {1}{3!} dx^{\m_1}\ldots
dx^{\m_{3}}\ep_{\m_1\ldots\m_4}g^{\m_4 \n}\6_\n\ph
\label{newch}\eeq
which shows that assumption \Gl{CPLiii} is not valid in this
case (as the reader may verify, $b^3_{inv}$ can indeed
be cast in the form \Gl{ch0a} and therefore
\Gl{newch} is equivalent to $\7F^I \7F^J
d_{IJ}\approx -s_1\be$, with $\be\in\cA_2$).

The physically most important changes in the results
if \Gl{CPLiii} is not satisfied are:
\ben
\item[(a)] Some BRST invariant
polynomials $P(q,f)$ which are
non-trivial if \Gl{CPLiii} holds become trivial. In particular
this can modify the results on chiral anomalies since some of
them can become trivial. In the above four-dimensional
example this happens to the abelian chiral candidate anomalies
given by $C^{abel}F^I F^Jd_{IJ}$ where
$C^{abel}$ is an abelian ghost. Namely \Gl{newch} implies
\beq C^{abel}F^I F^Jd_{IJ}= sb^{0,4}+db^{1,3}
\label{trivchiral}\eeq
where
\[ b^{0,4}=A^{abel}b^3_{inv}+d^4x C^{abel}\ph^*,\qd
b^{1,3}=C^{abel}b^3_{inv}-i_\xi b^{0,4}.  \]
Here $A^{abel}=dx^\m\A \m{abel}$ is the abelian connection 1-form
corresponding to $C^{abel}$,
$b^3_{inv}$ denotes the 3-form given in \Gl{newch}, $\ph^*$ is
the antifield of $\ph$ and
$i_\xi b^{0,4}$ is the interior product (contraction)
of the diffeomorphism ghost with the 4-form $b^{0,4}$
(explicitly one has $i_\xi=\xi^\m \6/\6(dx^\m)$ on local forms).
\item[(b)]
Any characteristic class of form degree $n$ which is trivial in
the equivariant characteristic cohomology
corresponds to a conserved Noether current
which, when written as an $(n-1)$-form, contains a Chern--Simons
$(2k-1)$-form.
Indeed, $P^{2k}(F):=d_{I_1\ldots I_k}F^{I_1}\ldots F^{I_k}\approx
db^{2k-1}_{inv}$ implies that $a^{2k-1}=q^{2k-1}(A,F)-
b^{2k-1}_{inv}$ is weakly closed ($da^{2k-1}\approx 0$)
where $q^{2k-1}(A,F)$ denotes the Chern--Simons form whose
exterior derivative equals $P^{2k}(F)$.
Hence, $a^{2k-1}$ indeed represents a conserved Noether current
in $n=2k$ dimensions.
In the above four-dimensional example the conserved Noether
current arising from \Gl{newch} is given by
\beq j^\m=eg^{\m\n}\6_\n \ph+e\ep^{\m\n\r\si}  d_{IJ}
(\A \n{I}\6_\r \A \si{J}+\s0 13\, \f KLJ
\A \n{I} \A \r{K} \A \si{L})
\label{except2}\eeq
and corresponds to the global symmetry of the action under
$\ph\rightarrow \ph+constant$. The latter is represented in
$H^{-1,n}(s|d)$ simply by the volume form $d^nx \ph^*$
(the non-trivial global symmetries and Noether currents
correspond one-to-one to the cohomology classes of $H^{-1,n}(s|d)$,
cf. \cite{bbh1} and section \ref{discussion}).
\een

The above example illustrates also that the
assumption \Gl{CPLiii} is
physically reasonable since couplings of
the type occurring in \Gl{cp4}
are not present in the standard
Einstein--Yang--Mills Lagrangian \Gl{xx1}.

\mysection{Interpretation and discussion of the results}\label{discussion}

\subsection{Characteristic cohomology}\label{charcoh}

We first summarize the implications of our results for the
characteristic cohomology, i.e. for the weak cohomology
of $d$ in the space of local  $p$-forms
which depend neither on the ghosts nor on antifields
($p<n$). As we have anticipated in the
introduction, there are two types of cohomology classes, denoted
by (i.a) and (i.b) respectively. Those
of the type (i.a) are represented by the forms
$\a_m^{0,k_m}$ introduced and discussed in section \ref{APL}.
They are closed even off-shell but fail to be exact
because of the non-trivial De Rham cohomology of the vielbein
manifold. Therefore we call them
topological conserved $p$-forms.

The cohomology classes of type (i.b) are represented
by local $p$-forms $a^{p}$ that are only weakly closed
and not weakly exact. They are therefore called
dynamically conserved $p$-forms and
can be obtained from the cohomological groups
$H^{p-n,n}(s|d)$.
Indeed, $da^p\approx 0$ is equivalent to $da^p=-\de \4a^{p+1}$
which implies $da^p=-sa^{p+1}$ for some
$(p+1)$-form $a^{p+1}=\4a^{p+1}+\ldots$ with
ghost number $-1$, cf. \cite{Henneaux3,bbh1}. Note that
$a^{p+1}$ vanishes only if $a^p$ is a
topological conserved $p$-form up to a $d$-exact form.
It is then straightforward, using the
arguments of section \ref{3}, to show that $a^p$ corresponds
via the descent equations indeed to a representative of a non-trivial
cohomology class of $H^{p-n,n}(s|d)$, unless
it is a topological conserved $p$-form up to a
$d$-exact form\footnote{Note:
these descent equations go down to a zero-form according to
section \ref{3} and thus $a^p$ is just the $p$-form with ghost
number 0 occurring in them.}.

The results of section \ref{volume} show that $H^{g,n}(s|d)$ is
empty for $g<-2$. This implies that
any weakly closed $p$-form is weakly exact for $p<n-2$, up
to a topological conserved $p$-form.
The cohomology classes of $H^{-2,n}(s|d)$ are represented
by the solutions $a^{-2,n}_4$ in the notation of
section \ref{volume} and are simply given by the volume forms
$d^nx C^*_\La$, independently
of the particular action. In contrast, the explicit form of
the corresponding dynamically conserved $(n-2)$-forms depends
on the chosen action. For the
standard Lagrangian \Gl{xx1} they are simply given by
\[ a^{n-2}_{\La}\simple\s0 {1}{2(n-2)!}(-)^{n+1} dx^{\m_1}\ldots
dx^{\m_{n-2}}\ep_{\m_1\ldots\m_n}F^{\m_{n-1}\m_n}_{\La}.
\]
Finally the dynamically conserved $(n-1)$-forms are just
provided by
the non-trivial Noether currents\footnote{In
this context a Noether current is called
trivial if it reduces on-shell to an identically conserved
current of the form $\6_\n S^{\n\m}$ with
$S^{\n\m}=-S^{\m\n}$. A global
symmetry is called trivial  if it reduces on-shell to a gauge transformation
with field dependent parameters, cf. \cite{bbh1} for details.}, up to
topological conserved $(n-1)$-forms.
In fact there are only two types of
cohomology classes of $H^{-1,n}(s|d)$ if
\Gl{CPLiii} holds. Their representatives
have been denoted in section \ref{volume}
by $a^{-1,n}_3$ and $a^{-1,n}_4$ respectively.
The latter occur only if ``free" abelian gauge fields are present.
The former correspond bijectively to the functions
$G^*_\De$ introduced in section \ref{notation} and
are given explicitly by
\[ a^{n-1}_{\De}= \s0 1{(n-1)!}\, dx^{\m_1}\ldots dx^{\m_{n-1}}
\ep_{\m_1\ldots \m_n}\E a{\m_n}J^a_\De(\cT).\]
The corresponding Noether currents satisfying $\6_\m j^\m_\De
\approx 0$ are given by
\beq j^\m_\De=e\E a{\m}J^a_\De(\cT).\label{noethercurr}\eeq

To summarize, we obtain almost the
same result for the characteristic cohomology as
in \cite{bbh1}. The only difference is the occurrence
of the topological conserved $p$-forms.
The reason why the non-contractibility
of the vielbein manifold does not lead to further modifications
may be seen from the proof of lemma \ref{lem7}.

The results on the equivariant characteristic cohomology have
been summarized already in section \ref{ass} and are not
repeated here.

\subsection{Covariance properties of the Noether currents and
of the global symmetries}

The results of subsection \ref{charcoh} allow us to conclude
that all non-trivial Noether currents
can be covariantized if ``free" abelian gauge fields are absent and
assumption \Gl{CPLiii} holds. Indeed, we have seen
that then all non-trivial
Noether currents can be assumed to be of the form
\Gl{noethercurr}. The latter are invariant under
local Lorentz and Yang--Mills
gauge transformations and transform as contravariant vector
densitites under general coordinate transformations
due to \Gl{r9a}, i.e. they satisfy
\beq sj^\m_\De=\6_\n(\xi^\n j^\m_\De)-(\6_\n \xi^\m)j^\n_\De\ .
\label{ch12}\eeq
The corresponding (infinitesimal) global symmetries of the action are
encoded in the part $X^*_\De$ of $G^*_{\De}$. Recall
that the $X^*_\De$ are defined only up to functions of the form
$\de M(\cT,\cT^*)+D_a N^a(\cT,\cT^*)$.
We can therefore remove all covariant derivatives
acting on the antifields by subtracting suitable parts
$D_a N^a$ from $X^*_\De$ and
assume with no loss of generality that
\beq X^*_\De=\7y^*_\Ly X_\De^\Ly(\cT)+
\7e_a^{*b}X_{\De b}^a(\cT) +\7A_i^{*a}X_{\De a}^i(\cT) .
\label{r9c}\eeq
The global symmetries of the classical action, denoted by
$\de_\De$, are then just given by
\beq \de_\De y^\Ly=X_\De^\Ly(\cT), \qd
\de_\De\e \m{a}= \e \m{b} X_{\De b}^a(\cT),\qd
\de_\De\A \m{i}= \e \m{a} X_{\De a}^i(\cT).\label{ch11}\eeq
Due to $\de_I X^*_\De=0$
these transformation are gauge-covariant in the
sense that $(\phi +\de_\De\phi)$ transforms exactly as
$\phi$ under diffeomorphisms, local Lorentz and Yang--Mills
transformations. Together with remark (b) at the end of section
\ref{volume} we conclude:
\begin{theorem}
If ``free" abelian gauge fields are absent and
\Gl{CPLiii} holds, then

(a) all non-trivial global symmetries
can be chosen to be gauge-covariant, i.e. to be of the form
\Gl{ch11} where the covariance properties of the $X$'s are
indicated by their indices $\Ly,a,b,i$;

(b) all non-trivial Noether currents can be
chosen to be gauge invariant under Lorentz and
Yang--Mills transformations
and to transform as contravariant vector-densities
under diffeomorphisms. More precisely,
they are of the form \Gl{noethercurr}
and satisfy \Gl{r9a} resp. \Gl{ch12}.

If \Gl{CPLiii} does not hold, then there are
non-trivial Noether currents
which can be chosen to be
of the form (b) only up to Chern--Simons terms. These
Noether currents correspond one-to-one to those characteristic
classes of form degree $n$
which become trivial in the equivariant characteristic cohomology.
\label{globalsymm}
\end{theorem}

We finally remark that
(a) and (b) do not always hold in presence of
``free" abelian gauge fields. This is easily verified
for the standard Lagrangian
\Gl{xx1} if the latter contains more than one ``free"
abelian gauge field. Namely then \Gl{xx1} has
the global symmetry $\de_\epsilon \A \m{\La}=
\epsilon_{\La\Lb}\A \m{\Lb}$ where $\epsilon_{\La\Lb}=
-\epsilon_{\Lb\La}$ are constants ($\de_\epsilon$
vanishes on all other fields). Evidently $\de_\epsilon$
is not gauge-covariant in the above sense (and cannot be
covariantized) and one easily checks that
the corresponding Noether current does not satisfy
\Gl{ch12} (the corresponding
cohomology class of $H^{-1,n}(s|d)$ is represented
by \Gl{n7}, choosing $P^2(\Th(C))=\s0 12\ep_{\La\Lb}C^\La C^\Lb$ there).

Theorem \ref{globalsymm} is analogous to the result
we have obtained in \cite{currents} for Yang--Mills theories (in flat
spacetime). Note however that in the gravitational case all
non-trivial global symmetries and Noether currents
can be chosen such that they do not depend explicitly on
the spacetime coordinates $x^\m$, contrary
to the results in flat spacetime.

\subsection{Consistent deformations of the action}

The BRST cohomology on local functionals with ghost number 0
provides the
possible consistent deformations of the action and its
gauge symmetries. A systematic construction of
such deformations has been described
in \cite{hen}. One looks
for a functional with ghost number 0 of the form
\beq\act^{ext}=\act^{(0)}+g\act^{(1)}+
g^2\act^{(2)}+\ldots\eeq
satisfying the master equation,
\beq \left(\act^{ext},\act^{ext}\right)=0\ .\label{Sext,Sext}\eeq
Here $\act^{(0)}$ denotes the original solution
of the classical master equation and $g$ is a ``coupling constant".
Expanding \Gl{Sext,Sext} in powers of $g$ one finds that
the $\act^{(k)}$ have to satisfy
\beq \left(\act^{(0)},\act^{(0)}\right)=0\ ,\qd
\left(\act^{(0)},\act^{(1)}\right)=0\ ,\qd
\left(\act^{(1)},\act^{(1)}\right)+
2\left(\act^{(0)},\act^{(2)}\right)=0\ ,\qd
\ldots\label{system}\eeq
where the first equation is just the original master equation
and thus satisfied by assumption. The second equation requires
$\act^{(1)}$ to be invariant under the original BRST operator $s$
whereas the third condition requires
$(\act^{(1)},\act^{(1)})$ to be $s$-exact. Contributions
to $\act^{(1)}$ which are already $s$-exact
can be neglected since they can be absorbed by
field redefinitions \cite{hen}.

Now, one can distinguish two types of deformations, denoted by
(ii.a) and (ii.b) respectively. Those of type (ii.a) modify
only the classical action but not the form of
its gauge symmetries. They correspond to antifield
independent local functionals $\act^{(1)}$ which are BRST closed.
For these, there is no higher order obstruction to the deformation since
$\left(\act^{(1)},\act^{(1)}\right)$ evidently vanishes. Thus, one
can set
$\act^{(i)}=0$ for $i>1$. Deformations
of the type (ii.b) modify simultaneously the
classical action and the form (and, in general,
the algebra) of its gauge symmetries. They correspond
to solutions
$\act^{(1)}$ of the first order condition
$\left(\act^{(0)},\act^{(1)}\right)=0$ that
depend non-trivially on the antifields.  Such first order deformation may
be obstructed at higher order, e.g. if
$\left(\act^{(1)},\act^{(1)}\right)$ is not BRST
exact in the space of local functionals.  These higher order
obstructions will not be discussed here.

We have seen above that $\act^{(1)}$ is non-trivially
$s$-invariant. Hence, it arises from the cohomology classes
of $H^{0,n}(s|d)$.
The results of section \ref{volume}
show that the latter are represented by volume forms
which do not depend on antifields at all, unless the
Yang--Mills gauge group contains abelian factors.
Indeed, the only antifield dependent ghost number 0-solutions
occurring in section \ref{volume} are the
$a_3^{0,n}$ and $a_4^{0,n}$ and
all of them involve abelian factors of $G_{YM}$.
We conclude that  deformations
of the type (ii.b) do not exist if $G_{YM}$ is semi-simple,
and that the only possible deformations are in that case
obtained simply by adding further gauge invariant pieces to the action
(involving e.g. higher powers in the curvatures or
higher order derivatives).

If abelian factors are present, there are in general
also deformations of type (ii.b). Their precise form
depends on the original classical action and its
global symmetries. We just note that among
these deformations there are those whose part
$\act^{(1)}$
includes the Yang--Mills cubic vertex for a
set of originally non-interacting abelian gauge fields
and the non-abelian extension of their
gauge transformations, cf. \cite{hen}
($\act^{(2)}$ then provides the Yang--Mills quartic vertex).

\subsection{Candidate anomalies}

The candidate anomalies correspond to the representatives
of the cohomology classes of $H^{1,n}(s|d)$. We distinguish
three types of them, denoted by (iii.a), (iii.b) and
(iii.c).

Those of type (iii.a) are given by the solutions
\Gl{top1} and \Gl{top2} with ghost number 1.
Since they stem from the non-trivial
De Rham cohomology of the vielbein manifold, we call them
topological candidate anomalies. Solutions \Gl{top1} with
ghost number 1 exist only
for $n=n_0+4k$ with $n_0=6,9,20,35$ ($k=0,1,\ldots$) as follows
from the Lie algebra cohomology of $so(n)$,
whereas the solutions \Gl{top2}
occur only in presence of ``free" abelian gauge fields.
An interpretation of the topological candidate anomalies
is not known to us.

The candidate anomalies of type (iii.b) and (iii.c) arise
solely from cohomology classes of $\cA_2$ with ghost number $(n+1)$.
Those of type (iii.b) do not involve the antifields. They
have been classified already in \cite{grav} where however
antifield dependent counterterms which can cancel
these candidate anomalies have not been considered.
Following \cite{grav}, to which we refer
for a thorough analysis, there are two types of
candidate anomalies (iii.b), those of the ``trace type" and the
``chiral anomalies" which in our notation are
the solutions $a^{1,n}_1$ and $a^{1,n}_2$ respectively, cf.
section \ref{volume}. The former are present only if the
gauge group contains abelian factors. Our results show that all
chiral candidate anomalies given in \cite{grav}
remain non-trivial even if one includes
antifield dependent counterterms, at least if
\Gl{CPLiii} is satisfied (remark (a) at the
end of section \ref{volume} shows that some chiral candidate
anomalies can become trivial if \Gl{CPLiii} is not valid).

The candidate anomalies (iii.c) depend non-trivially
on the antifields. They are denoted by
$a_3^{1,n}$ and $a_4^{1,n}$ in section \ref{volume}.
Again, these solutions occur only if the gauge group contains
abelian factors.

We conclude that the well-known chiral anomalies
are the only candidate anomalies if the Yang--Mills gauge
group is semi-simple, apart from the topological candidate
anomalies occurring in $n=6,9,10,13,\ldots$ spacetime dimensions.

\mysection{Acknowledgements}

This work has been supported in part by research funds from
F.N.R.S., by a research contract with the Commission of the
European Communities, and by the research council
(DOC) of the K.U. Leuven.

\appendix

\mysection{Conventions}\label{ap1}

Minkowski metric, $\ep$-tensor, volume forms:
\beann & &\et_{ab}=diag(1,-1,\ldots,-1),\\
& &\ep^{a_1\ldots a_n}=\ep^{[a_1\ldots a_n]},\qd
\ep^{0\ldots(n-1)}=1,\qd \ep_{\m_1\ldots \m_n}=
\ep_{a_1\ldots a_n}\e {\m_1}{a_1}\ldots\e {\m_n}{a_n},\\
& & d^nx=dx^{0}\ldots dx^{n-1}=
\s0 {(-)^{n+1}}{n!}dx^{\m_1}\ldots dx^{\m_n} \ep_{\m_1\ldots \m_n}/e .
\eeann
(Anti-) Symmetrization of indices, symmetrized trace of matrices:
\beann T_{(a_1\ldots a_k)}&=&\0 1{k!}\sum_{\pi}
                             T_{a_{\pi(1)}\ldots a_{\pi(k)}}\ ,\\
T_{[a_1\ldots a_k]}&=&\0 1{k!}\sum_{\pi}(-)^{sign(\pi)}
                             T_{a_{\pi(1)}\ldots a_{\pi(k)}}\ ,\\
Str(M_1\ldots M_k)&=&\0 1{k!}\sum_{\pi}
                         tr( M_{\pi(1)}\ldots M_{\pi(k)})
                                                              \eeann
where $\sum_\pi$ runs over all permutations $\pi$
in the symmetric group $S_k$.
\bsk

\noindent Antibracket:
\[ (X,Y)=\int d^nx\left( \0 {\de^RX}{\de\PH^A(x)}
\0 {\de^LY}{\de\PH^*_A(x)}
    -\0 {\de^RX}{\de\PH^*_A(x)}\0 {\de^LY}{\de\PH^A(x)}\right)\]
where $\de^R/\de Z(x)$ and $\de^L/\de Z(x)$
denote functional right- and left-derivatives.
The BRST operator $s$ is defined by means of the antibracket with
the proper solution $\act$ of the classical master equation through
\[ s\, X   = (\act,X).\]

\mysection{The Lie algebra cohomology of $\cG$ in $\cA_2$
}\label{ap2}

The Lie algebra
cohomology of $\cG$ in $\cA_2$ can be described in terms of its
primitive elements \Gl{r7} and the functions in $\ainv$ defined
in \Gl{defainv} \cite{LAC}:
\bea & &\gamma_{\cG}\a=0,\ \a\in \cA_2\qd\LRA\qd \a=\a^i\,
P_i(\Th(\7C))+\gamma_{\cG}\be,\
\a^i\in\ainv;\label{liecohom1}\\
& &\a^i\, P_i(\Th(\7C))=\gamma_{\cG}\be,\qd \a^i\in\ainv\qd
\LRA\qd \a^i=0\qd \forall\, i
\label{liecohom2}\eea
where $\be\in\cA_2$ and
the $P_i(\Th)$ are linearly independent polynomials in
the $\Th_\LAB$ and $\ainv$ is the
subspace of $\cA_2$ containing the $\de_I$-invariant
functions which do not depend on the ghosts $\7C^I$,
\beq \ainv=\{ \a\in\cA_2:\ \0 {\6\a}{\6\7C^I}=\de_I\a=0 \}.
\label{defainv}\eeq

The Lie algebra cohomology of $so(n-k,k)$ has in
$n=2r$ and $n=2r+1$ dimensions $r$
primitive (real) elements given by
\beann
\Th_{\LAB}(C)&=&N_{\LAB}
C_{a_1}{}^{a_2}B_{a_2}{}^{a_3}\ldots
B_{a_{2\LAB}}{}^{a_1},\qd \LAB=1,2,\ldots,r-1\ , \\
\Th_{r}(C)&=&N_r\left\{\ba{ll}
C_{a_1}{}^{a_2}B_{a_2}{}^{a_3}\ldots B_{a_{2r}}{}^{a_1} &
\mbox{for}\ n=2r+1 \\
\ep_{a_1 b_1\ldots a_r b_r}C^{a_1 b_1}
         B^{a_2 b_2}\ldots B^{a_r b_r} & \mbox{for}\ n=2r
\ea\right.
\eeann
where $B_a{}^b = C_a{}^cC_c{}^b$. These $\Th_{\LAB}$
agree for proper normalization factors $N_{\LAB}$ with \Gl{r7},
using $m_{\LAB}=2\LAB$ and choosing for $\{T^{\LAB}_{ab}\}$
the adjoint representation
of $so(n-k,k)$ except for the case $n=2r$, $\LAB=r$
in which one needs the spinor representation
\cite{orai}.

Analogously one constructs the $f_{\LAB}$ for $so(n-k,k)$
in the cases $n=2r$ and $n=2r+1$:
\beann
f_{\LAB}&=&
\7F_{a_1}{}^{a_2}\7F_{a_2}{}^{a_3}\ldots\7F_{a_{2\LAB}}{}^{a_1}
,\qd \LAB=1,2,\ldots,r-1\ ,\\
f_{r}&=&\left\{\ba{ll}
\7F_{a_1}{}^{a_2}\7F_{a_2}{}^{a_3}\ldots\7F_{a_{2r}}{}^{a_1}
& \mbox{for}\ n=2r+1 \\
\7F^{a_1b_1}\ldots\7F^{a_rb_r}\ep_{a_1b_1\ldots a_rb_r}
& \mbox{for}\ n=2r.
\ea\right.
\eeann
Notice that $f_{\LAB}$ vanishes for $\LAB>n/4$, except for $f_r$
in $n=2r$ dimensions.

\mysection{BRST invariant subalgebras of $\cA_2$}\label{ap4}

Since $\cA_2$ is a polynomial algebra and since $\cG=\cG_L+\cG_{YM}$
is reductive, we can uniquely decompose any
function $\a\in\cA_2$ into parts $\a^\la$ transforming under
$G$ according to a particular irreducible representation
$\la$. Hence, $\cA_2$
decomposes into subalgebras $\cA_2^\la$ containing only
functions transforming under $G$ according to $\la$.
Since $s$ commutes with all $\de_I$ it leaves invariant each
of these subalgebras, i.e. the image of $\cA_2^\la$ under $s$ is
contained in $\cA_2^\la$,
\[ s\, [\cA_2^\la]\subset \cA_2^\la.\]
This implies: (a) a function $\a\in\cA_2$ is $s$-invariant
iff all its parts $\a^\la$ are separately $s$-invariant and
(b) $\a\in\cA_2$ is $s$-exact in $\cA_2$ iff
the $\a^\la$ are $s$-exact in $\cA_2^\la$ respectively,
\beann
&\mbox{(a)} &
s\a=0,\qd \a\in\cA_2\qd\LRA\qd s\a^\lambda=0\qd \forall \lambda;\\
&\mbox{(b)} &
\a=s\be,\qd \a,\be\in\cA_2\qd\LRA\qd
\a^\lambda=s\be^\lambda\qd \forall \lambda.\eeann

\mysection{Cohomology of $\de$ in $\cA_2$}\label{ap3}

That $H_k(\de,\cA)$ is zero
 for positive antighost number $k$ is a standard
result and holds by construction of the Koszul--Tate complex,
see e.g. \cite{henteit}. Since functions in $\cA_1$ do not depend
on antifields at all, the subspaces of $\cA$ with non-vanishing
antighost number agree with those of $\cA_2$. Hence,
$H_k(\de,\cA_2)$ is zero for $k>0$.
According to appendix \ref{ap4},
$\cA_2$ can be decomposed into the subalgebras $\cA_{2}^\la$.
The latter can be further decomposed into subalgebras
$\cA_{2,q}^\la$ containing functions
with $\7C$-degree $q$. Now, $\de$ leaves invariant each
$\cA_{2,q}^\la$ since it does not change the $\7C$-degree and commutes
with all $\de_I$.
By the same reasoning as in appendix \ref{ap4}, we conclude
that $H_k(\de,\cA_{2,q}^\lambda)$ is zero in positive antighost number
$k$ for all $q$ and
$\lambda$ separately.

\mysection{Proof of the weak covariant Po\-in\-ca\-r\'e lem\-ma}\label{B}

We first prove that the
solution of \Gl{head0a} and \Gl{head0b}
is equivalent to the computation of the BRST cohomology in the
subspace $\ainv$ of $\cA_2$ defined in \Gl{defainv}.
\begin{lemma} Any solution $\a_\ell^i(\7\xi,\cT)$
of \Gl{head0a} and \Gl{head0b} with ghost number
$\ell$ can be completed to a function $\a\in\ainv$
representing a non-trivial cohomology class of
$H^\ell(s,\ainv)$. Conversely, any representative
of a non-trivial cohomology class of
$H^\ell(s,\ainv)$ contains a solution $\a_\ell^i(\7\xi,\cT)$ of
\Gl{head0a} and \Gl{head0b}.
\label{lem5}\end{lemma}

\proof{To prove the lemma one takes advantage of the
facts that  the BRST operator
reduces to $s_1+\de$ in $\ainv$ and that $s_1$ is nilpotent in $\ainv$,
\beq \a\in\ainv\qd\then\qd s_1^2\a=0,\qd s\a=(s_1+\de)\a,
\label{s-in-ainv}\eeq
which holds according to \Gl{m2}--\Gl{11} due to
$\gamma_\cG [\ainv]=s_2[\ainv]=0$. The proof is now straightforward
using standard arguments of
homological perturbation theory, cf. \cite{henteit}.
To see that any solution
of \Gl{head0a} can be completed to a BRST invariant function
$\a\in\ainv$ one acts with $s_1$ on \Gl{head0a}. Using \Gl{s-in-ainv}
this yields $\de(s_1\a_{\ell+1}^i)=0$ which implies
$\de\a_{\ell+2}^i+s_1\a_{\ell+1}^i=0$ for some
$\a_{\ell+2}^i\in\ainv$ due to the
acyclicity of $\de$ in positive antighost number, cf.
appendix \ref{ap3}. Iterating the
argument one concludes the existence of functions
$\a_p^i\in\ainv$ with $\7\xi$-degrees $p=\ell,\ell+1,\ldots, n$
such that $\a=\sum\a_p^i$
satisfies $(s_1+\de)\a=0$, and thus $s\a=0$ according
to \Gl{s-in-ainv}. Using \Gl{s-in-ainv} and
the acyclicity of $\de$ in positive antighost number
again, one shows analogously that
$\a\in\ainv$ is BRST exact in $\ainv$ iff
its antifield independent part is weakly $s_1$-exact.}

Theorem \ref{CPL} can therefore be proved by computing
$H^*(s,\ainv)$. To that end
we will take advantage of results derived in \cite{bbh1}
for the cohomology groups $H^*_*(\delta|d)$ which is the relative
cohomology of $\de$ modulo $d$ in the space of local forms.
However we cannot directly
adopt these results for the theories at hand since we have
assumed in \cite{bbh1} the field-antifield manifold to be
contractible to a point which fails to be true here because
of the non-contractibility of the vielbein manifold, cf. section
\ref{APL}. In particular the proof given in
\cite{bbh1} for normal theories uses a perturbative argument
which applies directly to gravitational theories only if one
splits the vielbein
into a constant background vielbein and a deviation $h_\mu^a$ from
it. The proof given in \cite{bbh1} works then only
in the space of forms which are polynomial in the derivatives of all
fields, but involve in general
formal infinite power series' in the undifferentiated
$h_\mu^a$, without guaranteeing that these series' are really
regular (smooth) in the $\e \mu{a}$ (for $\det(\e \mu{a})\neq 0)$.
We will first show now that the series' can be really assumed
to be regular in all cases we need. This is encoded in the following
lemma, as becomes clear from its proof:

\begin{lemma}
$H^G(s,\cA_2)$ is zero for $G<0$ and $0<G<n-2$,
$H^0(s,\cA_2)$ contains only one non-trivial cohomology class
represented by a constant, and the
non-trivial cohomology classes of $H^{n-2}(s,\cA_2)$ are
represented by the $q^*_{\La}$,
\bea \lefteqn{s\, \a=0,\qd \a\in\cA_2,\qd \gh{\a}=G\leq n-2}\nn\\
&\LRA&
\a=\de_G^{n-2} \la^{\La} q^*_{\La}+
\de_G^{0}\,\la+s\, \be,\qd \be\in\cA_2\label{o6}\eea
where $\la^{\La}$ and $\la$ are constants and
$\de_G^{p}$ denotes the Kronecker symbol
($\de_G^p=1$ if $G=p$ and $\de_G^p=0$ if $G\neq p$).
\label{lem7}\end{lemma}

\proof{Assume that $\a\in\cA_2$ is BRST invariant and has
ghost number $G<n-2$ (the case $G=n-2$ is considered separately below).
According to theorem
\ref{correspondence}, it corresponds via the descent equations
to a volume form $a^{G-n,n}$ solving \Gl{i1} or
it is BRST exact up to a constant. A constant can occur of course
only in the case $G=0$. Note that $a^{G-n,n}$ has ghost number $G-n<-2$.
General results on the local BRST cohomology derived first in
\cite{Henneaux3} and refined for normal theories in
\cite{bbh1,bbh2} state that $H^{g,n}(s|d)$ is
isomorphic to $H^n_{-g}(\de|d)$ for $g<0$, the latter
being the cohomology of $\de$ modulo $d$ in the space of local
$n$-forms with antighost number $(-g)$.

Now, in \cite{bbh1} we have proved that
the cohomology groups $H^n_{k}(\de|d)$ vanish for $k>2$ in normal
irreducible gauge theories treated there. Assume that the statement
remains true even in our case for the volume forms $a^{G-n,n}$ which
arise from BRST invariant functions $\a\in\cA_2$.
Then we could conclude that $a^{G-n,n}$ is trivial and thus that
$\a$ is BRST exact up to a constant which would prove the theorem
for $G<n-2$.

However, as mentioned above, the
proof given in \cite{bbh1} applies to the gravitational theories
in question directly only for
forms that are formal infinite power series' in the undifferentiated
$ \e \mu{a}$ (but still polynomials
in the derivatives of all fields, including $\e \m{a}$). Let us
call such forms ``quasi-local". Hence, the results
of \cite{bbh1} allow us only to conclude that $\a$ is $s$-exact in
the space of quasi-local functions, i.e. that
\beq \a=s\be+\de_G^{0}\,\la\label{ab0}\eeq
holds for some quasi-local function $\be$. We have to show now
that we can even assume $\be$ to be (strictly) local.
This becomes obvious when one expresses $\be$ in terms
of the generators introduced in section \ref{basis}
and decomposes it into parts $\be^{(r)}$ with definite degree
$r$ in the generators $U_l$ and $V_l$.
Due to \Gl{trivpair} and \Gl{tensorproduct2}
the counting operator for all the
$U$'s and $V$'s commutes with $s$. Therefore
\Gl{ab0} requires $s\be^{(r)}=0$ for all $r\neq 0$ and
$s\be^{(0)}=\a-\de_G^{0}\la$ since $\a$ does not depend on the
$U$'s and $V$'s at all due to $\a\in\cA_2$. Note that $\be^{(0)}$
is local since $\be$ is quasi-local. We conclude
$\be^{(0)}\in\cA_2$ which proves the lemma for $G<n-2$.

For $G=n-2$ the results of \cite{bbh1} (section 13)
and equation \Gl{sc*} show that
$a^{G-n,n}=a^{-2,n}$ is given by a linear
combination of the $n$-forms $d^nxC^*_{\La}$, up to a
piece which is trivial in the space of quasi-local forms.
The former corresponds via the descent equations to a linear combination
of the functions $q^*_\La$, cf. subsection \ref{notation}, i.e.
we conclude $\a-\la^\La q^*_\La=s\be$ for some quasi-local
function $\be$. Due to $q^*_\La\in\cA_2$ we can use now the
same argument as above to conclude $\a-\la^\La q^*_\La=s\be^{(0)}$
for $\be^{(0)}\in\cA_2$.

This completes the demonstration of
the lemma and shows simultaneously that the
results of \cite{bbh1} on
$H^p_k(\de|d)$ remain valid in our case for forms
corresponding via the descent equations
to BRST invariant functions $\a\in\cA_2$ (and for $p\leq n+k-2$).
Note that an analogous conclusion is in fact not true for
$\a\in\cA$. Indeed, the vielbein manifold enters through $\cA_1$
and gives rise to the functions $\a_m^{k_m,0}$
(and the forms $\a_m^{0,k_m}$) which represent
non-trivial cohomology classes of $H^{k_m}(s)$
($H^{k_m}(d)$) although they are $s$-exact ($d$-exact)
in the space of quasi-local functions (forms). Note
that this implies that any function \Gl{expansion2}
is $s$-exact in the space of quasi-local functions
except for those which involve
the constant representative of $H^0(s,\cA_1)$.}
\bsk

It is important to realize that to compute $H^*(s,\ainv)$
it is not sufficient to
determine only those BRST invariant functions $\a\in\ainv$
which are not BRST exact. Namely it can (and does)
happen that a function $\a\in\ainv$ is BRST exact in $\cA_2$
but not in $\ainv$. We shall determine first
these functions by means of lemma \ref{lem7}. To this end
we assume that $\a\in\ainv$ is BRST exact (in $\cA_2$) and
has ghost number $<n$,
\beq \a=s\, \be,\qd \a\in\ainv,\qd\be\in\cA_2,\qd \gh{\a}=G<n.
\label{o7}\eeq
We now differentiate \Gl{o7} with respect to $\7C^I$.
Since $\a$ does not depend on $\7C^I$, we obtain
\beq 0=(s\7\6_I+\7\6_I s)\be-s\, \7\6_I\be\label{zusatz}\eeq
where we have introduced the notation
\beq \7\6_I=\0 \6{\6C^I}\ .\label{co6b}\eeq
We now take advantage of the fact that $s\7\6_I+\7\6_I s$
represents $\de_I$ on $\cA_2$, as follows immediately from
\Gl{m2}--\Gl{s2}:
\beq \de_I=  s\7\6_I+\7\6_I s \ .   \label{co6c}\eeq
According to appendix \ref{ap4} we can assume $\de_I\be=0$
with no loss of generality due to $\de_I\a=0$. Hence, we
conclude from \Gl{zusatz} and \Gl{co6c}
\beq s\, (\7\6_I\be)=0.\label{o8}\eeq
For later use we note that the functions
$\7\6_I\be$ transform under the co-adjoint representation of $\cG$
due to $\de_I\be=0$:
\beq
\de_I\, (\7\6_J\be)=\f IJK\7\6_K\be.\label{o9}\eeq
This follows directly from $\de_I\be=0$ due to
\beq [\de_I,\7\6_J]=\f IJK\7\6_K.\label{zusatz2}\eeq
Note that the $\7\6_I\be$
have ghost number $G-2<n-2$ due to $\gh{\be}=\gh{\a}-1$.
Therefore we can conclude from \Gl{o8}, using lemma \ref{lem7}:
\beq\7\6_I\be=d_I\, \de_G^2+s\, \be_I,\qd d_I=const.\label{o10}\eeq
The Kronecker-symbol $\de_G^2$ indicates
that constants can contribute to $\7\6_I\be$ only if the latter
have vanishing ghost number. Recall that the $\7\6_I\be$ transform
according to \Gl{o9} under $\cG$. This implies that both $s\be_I$ and
$d_I$ can be assumed to transform under the co-adjoint representation of
$\cG$ since the decomposition
$d_I+s\, \be_I$ is unique, as the constants are not
BRST exact\footnote{If they were then the theory would be inconsistent
since the equations of motion would imply ``$5\approx 0$".}.

On the one hand this implies that $d_I$ can be
different from zero only if its index $I$ refers to an
abelian element of $\cG$ since the semi-simple part of $\cG$
does not admit invariant tensors of the type $d_I$,
\[ \f IJK d_K=0.\]
On the other hand
we conclude, using appendix \ref{ap4} again,
that the functions $\be_I$ can be assumed to transform under $\cG$
according to the co-adjoint representation since this
holds already for $s\be_I$:
\beq\de_I\, \be_J=\f IJK\be_K\label{o11}\eeq
We now differentiate \Gl{o10} with respect to $\7C^J$. This yields,
using \Gl{co6c} and \Gl{o11},
\beq  \7\6_J\7\6_I\be= \f JIK\be_K-s\, (\7\6_J\be_I).\label{o12}\eeq
Recall that $\7\6_I$ are derivatives with respect to anticommuting
generators and that the structure constants $\f IJK$ are antisymmetric:
\beq \7\6_I\7\6_J=-\7\6_J\7\6_I,\qd \f IJK=-\f JIK.\label{o13}\eeq
Hence, \Gl{o12} implies $s\, \7\6_{(J}\be_{I)}=0$. Due
to $\gh{\7\6_J\be_I}=G-4<n-4$ we conclude, using lemma
\ref{lem7} again:
\beq\7\6_{(I}\be_{J)}=d_{IJ}\, \de_G^4+s\, \be_{IJ},
\qd d_{IJ}=const.\label{o14}\eeq
By means of arguments used already above we
conclude that the $d_{IJ}$ are the components of a symmetric invariant
tensor of $\cG$,
\[ d_{IJ}=d_{JI},\qd \f IJL d_{LK}+\f IKL d_{JL}=0\ , \]
and that the $\be_{IJ}$ can be assumed to be symmetric too
and to transform according to their index picture under $\cG$:
\[ \be_{IJ}=\be_{JI},\qd
\de_I\, \be_{JK}=\f IJL\be_{LK}+\f IKL\be_{JL}.\]
Iterating this procedure one concludes
the existence of a chain of functions $\be_{I_1\ldots I_m}$,
$m=0,1,\ldots$ satisfying
\bea \7\6_{(I_1}\be_{I_2\ldots I_m)}=
d_{I_1\ldots I_m}\, \de_G^{2m}+s\, \be_{I_1\ldots I_m},\qd
\label{o16}\eea
where $d_{I_1\ldots I_m}$ are the components of
invariant symmetric tensors of $\cG$,
\beann & &d_{I_1\ldots I_m}=d_{(I_1\ldots I_m)}=const. ,\\
& &\sum_{r=1}^m\f I{I_r}J
d_{I_1\ldots I_{r-1}JI_{r+1}\ldots I_m}=0\eeann
and the $\be_{I_1\ldots I_m}$ can be assumed to satisfy
\beann & & \be_{I_1\ldots I_m}=\be_{(I_1\ldots I_m)},\qd
\gh{\be_{I_1\ldots I_m}}=G-2m-1,\\
& & \de_I\, \be_{I_1\ldots I_m}=\sum_{r=1}^m\f I{I_r}J
\be_{I_1\ldots I_{r-1}JI_{r+1}\ldots I_m}.\eeann
The occurrence of the constants $d_{I_1\ldots I_m}$ implies
the presence of functions which are
BRST exact in $\cA_2$ but possibly not in $\ainv$. These functions are
invariant polynomials in the $\7F^I$:
\beq P_{2m}(\7F)=d_{I_1\ldots I_m}\7F^{I_1}\ldots\7F^{I_m}.\label{o18}\eeq
This follows directly from \Gl{o7} and \Gl{o16}. To show this, we
first decompose the functions $\be_{I_1\ldots I_m}$
into parts with definite $\7C$-degree:
\[ \be_{I_1\ldots I_m}=\sum_{k\geq 0}\be^k{}_{I_1\ldots I_m},\qd
\7C^I\0 \6{\6\7C^I}\be^k{}_{I_1\ldots I_m}=k\be^k{}_{I_1\ldots I_m}.\]
Taking advantage of \Gl{s0}--\Gl{s2}, we conclude from \Gl{o7}
\beq \a=(\de+s_1)\be^0+s_2\be^1=s\be^0+\7F^I\7\6_I\be^1 \label{o19}\eeq
since $(\de+s_1)\be^0+s_2\be^1$ are the only contributions to $s\be$
with vanishing $\7C$-degree
and since $(\de+s_1)\beta^0=s\beta^0$ holds due to
$\beta^0\in\ainv$. The last term in \Gl{o19} is now
replaced by the following expression obtained from \Gl{o10}:
\bea \7F^I\7\6_I\be^1&=&
\7F^I(d_I\de_G^2+(\de+s_1)\be^0{}_I+s_2\be^1{}_I)\nn\\
&=&\de_G^2\, P_2(\7F)+s(\7F^I\be^0{}_I)+
\7F^I\7F^J\7\6_{(J}\be^1{}_{I)} \label{o20a}\eea
where we used $(\de+s_1) \7F^I=0$, $\7F^I\be^0{}_I\in\ainv$
and \Gl{s-in-ainv}, and $\7F^I\7F^J\7\6_{J}\be^1{}_{I}
=\7F^I\7F^J\7\6_{(J}\be^1{}_{I)}$ (the latter holds since
the $\7F^I$ are commuting quantities). The last term in \Gl{o20a}
is now replaced by an expression following from \Gl{o14} and the
procedure is iterated until the number of occurring
$\7F$'s exceeds $n/2$
(recall that monomials $\7F^{I_1}\ldots\7F^{I_k}$ vanish for $k>n/2$
since $\7F^I$ has $\7\xi$-degree 2). This results finally in
\beq \a=\de_G^{2m}\, P_{2m}(\7F)+s\, \7\be,\qd
     \7\be\in\ainv\label{o20}\eeq
where
\[ \7\be=\sum_{k\leq n/2}
\7F^{I_1}\ldots\7F^{I_k}\be^0{}_{I_1\ldots I_k}.\]
We have thus proved that the
invariant polynomials in the $\7F^I$ are the only
functions with ghost number $<n$ which are BRST exact in
$\cA_2$ but possibly
not BRST exact in $\ainv$. Together with
Lemma \ref{lem7} this shows that any BRST invariant function
in $\ainv$ with ghost number $G<n-1$ is BRST exact
in $\ainv$ up to an invariant polynomial in the $\7F^I$ (including
a constant for $G=0$) and a linear combination of those $q^*_\La$ which
are in $\ainv$. The latter are just the $q^*_{\La^0}$ occurring
in theorem \ref{CPL}.

We still have to consider the cases
$G=n-1$ and $G=n$. This is very easy. Namely any
function $\a\in\ainv$ with ghost number $n$ is necessarily
BRST invariant since $n$ is the maximal ghost number a
function in $\ainv$ can have. Furthermore,
by definition of $\ainv$, $\a$ has
the form $\a=\cE L(\cT)$ with $\de_I L(\cT)=0$
and $\cE$ as in \Gl{vol}.

The most general form of a
function $\a\in\ainv$ with ghost number
$(n-1)$ is $\a= X^*(\cT,\cT^*)\cE+
J^a(\cT)\6\cE/\6\7\xi^a$ where
$X^*(\cT,\cT^*)$ is $\de_I$-invariant and has
antighost number 1, and the $J^a$ satisfy \Gl{r9a}.
$s\a=0$ then requires $X^*$ and $J^a$ to solve
\Gl{r9} and therefore
$X^*$ is a linear combination of the $X^*_\De$ up
to a piece $\de M(\cT,\cT^*)+D_a N^a(\cT,\cT^*)$,
cf. subsection
\ref{notation}. This implies immediately
that $\a$ is a linear combination of the $G^*_\De$
up to a piece which is $s$-exact in $\ainv$
(and given by $s[M\cE-(-)^{\ep(N^a)} N^a\6\cE/\6\7\xi^a]$)
and a piece which does not
depend on antifields at all. Since $\a$ has ghost number $(n-1)$,
the latter is trivial in $\cA_2$ by corollary
\ref{cor1} and therefore it is trivial in $\ainv$
up to an invariant polynomial in the $\7F^I$ by the above result.
We conclude:
\begin{lemma}
Any BRST invariant function $\a\in\ainv$ is
BRST exact in $\ainv$ up to an invariant polynomial
in the $\7F^I$, a function of maximal $\7\xi$-degree
and a linear combination of the $G^*_\De$ and the
$q^*_{\La^0}$,
\beq s\a=0,\ \a\in\ainv\qd\LRA\qd
\a=P(\7F)+\la^{\La^0}q^*_{\La^0}+
\la^{\De}G^*_{\De} +\cE L(\cT)+s\be
\label{CPLI}
\eeq
where
\[ \de_IP(\7F)=\de_I L(\cT)=0, \qd\be\in\ainv.\]
\end{lemma}
This result implies immediately part (i) of theorem \ref{CPL}
because of lemma \ref{lem5}.

To prove part (ii) of the theorem we assume that $P(\7F^I)$ is
an invariant polynomial in the $\7F^I$ of the form
\Gl{o18} and is BRST exact in $\ainv$,
\beq P_{2m}(\7F)=s\a,\qd \a\in\ainv.
\label{pr3}\eeq
Now, from \Gl{sq} and \Gl{invariants} we know that
$P_{2m}(\7F)$ can also be written as the BRST variation
of a polynomial $\cP$ in the $q_\LAB$ and $f_\LAB$,
\beq
P_{2m}(\7F)= s \cP(q,f).\label{pr1}\eeq
We have to prove that $P_{2m}(\7F)$ vanishes for $2m<n-1$.
To that end we assume
$P_{2m}(\7F)\neq 0$ and show that this leads to a
contradiction for $2m<n-1$. First we note that $P_{2m}(\7F)\neq 0$
implies that the polynomial $\cP(q,f)$ occurring
in \Gl{pr1} has level $\leq 2m-2$,
\beq \cP(q,f)=\sum_{p\geq p_0}\cP_p,\qd N_{\7\xi}\cP_p
=p\cP_p,\qd p_0\leq 2m-2\label{pr2}\eeq
where $\cP_{p_0}\neq 0$ is understood.
Note that the level of a function in $\ainv$ cannot be smaller
than its ghost number. Since $\a$
has ghost number $(2m-1)$ we know that it takes the form
\beq \a=\sum_{p\geq 2m-1}\a_p,\qd N_{\7\xi}\a_p=p\a_p.\label{pr0}\eeq
{}From \Gl{pr1} and \Gl{pr3} we conclude
\beq s\, [\a-\cP(q,f)]=0.\label{pr4}\eeq
Consider now the case $2m<n-1$. Then $\a-\cP(q,f)$ has ghost number
$2m-1<n-2$ and therefore is BRST exact (in $\cA_2$)
according to lemma \ref{lem7},
\beq
2m<n-1:\qd
\a-\cP(q,f)=s\, \be.\label{pr5}\eeq
A constant cannot occur here because $\a-\cP(q,f)$ has
odd ghost number (given by $2m-1$).
\Gl{pr5} is now analyzed analogously to \Gl{comp3}.
Indeed, \Gl{pr5} contains a set of equations
\Gl{przz7}--\Gl{przz10} to which $\a$ does not contribute since
its level exceeds the level $p_0$ of
$\cP(q,f)$, cf. \Gl{pr2} and \Gl{pr0}. Using the same arguments
as in the discussion of \Gl{przz7}--\Gl{przz10}, we conclude
\beq \be=\7\be+{\cal Q}(q,f)+{\cal Q}'(q,f)+\ldots\eeq
where $\7\be$ satisfies
\beq \a-\7\cP(q,f)=s\, \7\be,
\qd \ell(\7\be)\geq \7p_0-1.
\label{pr13}\eeq
Here $\ell(\7\be)$ and $\7p_0$ are the levels of $\7\be$
and $\7\cP(q,f)$ respectively. The same arguments which we used
to derive \Gl{przz17}
lead now to the conclusion
\beq P^i_{\7p_0}(\7F)=s_1\, \7\be^i_{\7p_0-1}+\de\, \7\be^i_{\7p_0}
,\qd \7\be^i_{\7p_0-1}, \7\be^i_{\7p_0}\in\ainv\ .\label{pr17}\eeq
Acting with $s_1$ on this equation results in $\de(s_1\7\be^i_{\7p_0})=0$
due to $s_1 P^i_{\7p_0}(\7F) =0$ and \Gl{s-in-ainv}. Using the acyclicity
of $\de$ in positive antighost number (appendix \ref{ap3}), we conclude
$s_1\7\be^i_{\7p_0}+\de\7\be^i_{\7p_0+1}=0$
for some $\7\be^i_{\7p_0+1}\in\ainv$.
An iteration of the arguments (cf. proof of lemma \ref{lem5}) implies
the existence of $\7\be^i=\sum_{p\geq \7p_0-1}\7\be^i_p$ such that
\beq P^i_{\7p_0}(\7F)=s\, \7\be^i,\qd \7\be^i\in\ainv.\label{pr18}\eeq
This is an equation like \Gl{pr3}, but for lower ghost number since
we have $\7p_0<2m$. One can now repeat the arguments until the ghost
number drops to zero. This provides the desired
contradiction since a constant is not BRST exact.

Finally, consider \Gl{pr3} for
$2m=n-1$. We conclude from \Gl{pr4} by
means of lemma \ref{lem7} that $\a-\cP(q,f)$ is BRST exact up to
a linear combination of the $q^*_{\La}$:
\beq 2m=n-1:\qd
\a-\cP(q,f)=\la^{\La}q^*_{\La}+s\, \be,\qd
\be\in\cA_2,\qd \de_I\be=0.\label{pr19}\eeq
Suppose that all $q^*_{\La}$ are in
$\ainv$, i.e. that $\{q^*_{\La}\}=\{q^*_{\La^0}\}$. Then \Gl{pr19}
agrees with \Gl{pr5} if we replace there $\a$ by
$\a'=\a-\la^{\La}q^*_{\La}$. Hence, we
can adopt the analysis following \Gl{pr5} even in the case $2m=n-1$,
and conclude $P_{n-1}(\7F)=0$ if $\{q^*_{\La}\}=\{q^*_{\La^0}\}$.
This proves
\begin{lemma}
No non-vanishing $\de_I$-invariant polynomial in the $\7F^I$ of ghost
number $<n-1$ is BRST exact in $\ainv$,
\beq d_{I_1\ldots I_k}\7F^{I_1}\ldots \7F^{I_k}=
s\a,\ \a\in\ainv,
\ 2k<n-1\qd\LRA\qd d_{I_1\ldots I_k}=0.\label{lasteq}\eeq
\Gl{lasteq} remains valid for $2k=n-1$ if
$\{q^*_{\La}\}=\{q^*_{\La^0}\}$.
\end{lemma}
Part (ii) of theorem \ref{CPL} is equivalent to this lemma through
lemma \ref{lem5}.


\begin{thebibliography}{60}
\bibitem{Bryant}
R.L. Bryant and P.A. Griffiths, {\em Characteristic
Cohomology of Differential
Systems (I): General Theory}, Duke University Mathematics Preprint
Series, volume 1993 n$^0$1
(January 1993).
\bibitem{Wheeler}
C.W. Misner and J.A. Wheeler, {\em Ann. Phys.}  {\bf 2} (1957) 525.
\bibitem{Unruh}
W.G. Unruh, {\em Gen. Relativ. Gravit.} {\bf 2} (1971) 27.
\bibitem{hen}
G. Barnich and M. Henneaux, {\em Phys. Lett.} {\bf B311} (1993) 123.
\bibitem{Wald}
R. Wald, {\em Phys. Rev.} {\bf D33} (1986) 3613.
\bibitem{gravlit}
L. Alvarez-Gaum\'e and E. Witten, {\em Nucl. Phys.} {\bf B234} (1984) 269;
W. Bardeen and B. Zumino, {\em Nucl. Phys.} {\bf B244} (1984) 421;
F. Langouche, T. Sch\"ucker and R. Stora, {\em Phys. Lett.}
{\bf B145} (1984) 342;
L. Alvarez-Gaum\'e and P. Ginsparg, {\em Ann. Phys.} {\bf 161} (1985) 423;
L. Bonora, P. Pasti and M. Tonin, {\em J. Math. Phys.} {\bf 27} (1986) 2259;
T. Sch\"ucker, {\em Commun. Math. Phys.} {\bf 109} (1987) 167.
\bibitem{grav}
F. Brandt, N. Dragon and M. Kreuzer, {\em Nucl. Phys.} {\bf B340} (1990)
187.
\bibitem{brs}
C. Becchi, A. Rouet and R. Stora, {\em Commun.
Math. Phys.} {\bf 42} (1975) 127; {\em Ann. Phys.}
 {\bf 98} (1976) 287;
I.V. Tyutin, {\em Gauge invariance in field theory
and statistical mechanics},
Lebedev preprint FIAN,  n$^0$39 (1975).
\bibitem{brs2}
J. Zinn-Justin, {\em Renormalisation of gauge theories} in
{\em Trends in elementary particle theory}, Lecture notes in Physics
n$^0$37, Springer (Berlin: 1975).
\bibitem{brs3}
R.E. Kallosh, {\em Nucl. Phys.} {\bf B141} (1978) 141.
\bibitem{brs4}
B. de Wit and J.W. van Holten, {\em Phys. Lett.} {\bf B79} (1978) 389.
\bibitem{bv} I.A. Batalin and G.A. Vilkovisky, {\em Phys. Lett.}
{\bf B102} (1981) 27; {\em Phys. Rev.} {\bf D28} (1983) 2567;
{\em Phys. Rev.} {\bf D30} (1984) 508.
\bibitem{wz}
J. Wess and B. Zumino, {\em Phys. Lett.} {\bf B37} (1971) 95.
\bibitem{bbh1}
G. Barnich, F. Brandt and M. Henneaux,
{\em Local BRST cohomology in the antifield formalism: I. General
theorems},
preprint ULB-TH-94/06, NIKHEF-H 94-13, hep-th/9405109, to appear
in {\em Commun. Math. Phys.}
\bibitem{rapid}
G. Barnich, F. Brandt and M. Henneaux,
{\em Phys. Rev.} {\bf D51} (1995) R1435.
\bibitem{Fink} D. Finkelstein and C.W. Misner,
{\em Ann. Phys.} {\bf 6} (1959) 230.
\bibitem{Noether}
E. Noether, {\em Nachr. v. d. Kgl. Ges. d. Wiss. z. G\"ottingen,
Math.-phys. Kl.}, {\bf 2} (1918) 235.
\bibitem{anto}
C.G. Torre and I.M. Anderson, {\em Phys. Rev. Lett} {\bf 70} (1993) 3525.
\bibitem{Brandt} F. Brandt,
{\em Structure of BRS-invariant local functionals},
preprint NIKHEF-H 93-21, hep-th/9310123.
\bibitem{henfisch} J.M.L. Fisch and M. Henneaux, {\em Commun. Math.
Phys.} {\bf 128} (1990) 627;
M. Henneaux, {\em Nucl. Phys. B (Proc. Suppl.)} {\bf 18A} (1990) 47.
\bibitem{schweda} O. Moritsch and M. Schweda, {\em Helv. Phys. Acta}
{\bf 67} (1994) 289.
\bibitem{Takens} F. Takens, {\em J. Differential
Geometry} {\bf 14} (1979) 543; I.M. Anderson and
T. Duchamp, {\em Amer. J. Math.} {\bf 102} (1980) 781.
\bibitem{Anderson}
I.M. Anderson, {\em Contemp. Math.} {\bf 132} (1992) 51,
{\em The variational bicomplex},
Academic Press (Boston: 1994).
\bibitem{various}
A.M. Vinogradov, {\em  Sov. Math. Dokl.} {\bf
18} (1977) 1200,
{\bf 19} (1978) 144, {\bf 19} (1978) 1220;
M. De Wilde, {\em Lett. Math. Phys.} {\bf 5} (1981) 351; W.M.
Tulczyjew,
{\em Lecture Notes in Math.} {\bf 836} (1980) 22; P. Dedecker and
W.M. Tulczyjew, {\em
Lecture Notes in Math.} {\bf 836} (1980) 498; T. Tsujishita, {\em
Osaka J.of Math.} {\bf 19}
(1982) 311;
L. Bonora and P. Cotta-Ramusino, {\em Commun. Math. Phys.}
{\bf 87} (1983) 589;
P.J. Olver,
{\em Applications of Lie Groups to Differential Equations},
Graduate Texts in Mathematics, volume 107, Springer Verlag (New York:
1986);
R.M. Wald, {\em J. Math. Phys.} {\bf 31} (1990)
2378;
L.A. Dickey, {\em Contemp. Math.} {\bf 132} (1992) 307.
\bibitem{com}
F. Brandt, N. Dragon and M. Kreuzer,
{\em Nucl. Phys.} {\bf B332} (1990) 224;
M. Dubois-Violette, M. Henneaux, M. Talon and
C.M. Viallet, {\em Phys. Lett.} {\bf B267} (1991) 81.
\bibitem{Torre} C.G. Torre,
{\em Some remarks on gravitational analogs of magnetic charge},
preprint, gr-qc/9411014.
\bibitem{StoraZumino}
R. Stora, {\em Continuum Gauge Theories} in
{\em New Developments in Quantum Field Theory and Statistical Mechanics},
eds. M. L\'{e}vy and P. Mitter (Plenum: 1977), {\em Algebraic
Structure and Topological Origin of Anomalies} in {\em Progress in
Gauge Field Theory}, eds. G. 't Hooft et al. (Plenum: 1984);
B. Zumino, {\em Chiral
Anomalies and Differential Geometry} in
{\em Relativity, Groups and Topology II}, eds. B.S. De Witt and R. Stora
(North-Holland: 1984);
L. Baulieu, {\em Phys. Rep.} {\bf 129} (1985) 1.
\bibitem{sorella} S.P. Sorella, {\em Commun. Math. Phys.}
{\bf 157} (1993) 231.
\bibitem{Baulieu}
L. Baulieu and M. Bellon, {\em Nucl. Phys.} {\bf B266} (1986) 75;
L. Baulieu, M. Bellon and R. Grimm, {\em Nucl. Phys.} {\bf B294} (1987) 279;
F. Brandt, {\em Class. Quantum Grav.} {\bf 11} (1994) 849.
\bibitem{orai}
L. O'Raifeartaigh, {\em Group structure of gauge theories},
Cambridge University Press (Cambridge: 1986), section 5.6.
\bibitem{tal}
M. Dubois-Violette, M. Talon and C.M. Viallet,
{\em Commun. Math. Phys.} {\bf 102} (1985) 105.
\bibitem{lie}
F. Brandt, N. Dragon and M. Kreuzer,
{\em Nucl. Phys.} {\bf B332} (1990) 250.
\bibitem{Gilkey}
P. Gilkey, {\em Adv. in Math.} {\bf 28} (1978) 1.
\bibitem{bbh2}
G. Barnich, F. Brandt and M. Henneaux,
{\em Local BRST cohomology in the antifield formalism: II.
Application to Yang--Mills theory},
preprint ULB-TH-94/07, NIKHEF-H 94-15, hep-th/9405194, to appear
in {\em Commun. Math. Phys.}.
\bibitem{Henneaux3} M. Henneaux, {\em Commun. Math. Phys.} {\bf
140} (1991) 1.
\bibitem{currents}
G. Barnich, F. Brandt and M. Henneaux, {\em Phys. Lett.}
{\bf B346} (1995) 81.
\bibitem{LAC}
J.L. Koszul, {\em Bull.
Soc. Math. France} {\bf 78} (1950) 65;
C. Chevalley and S. Eilenberg, {\em Trans.
Am. Math. Soc.} {\bf 63} (1953) 589; G. Hochschild and J.P. Serre,
{\em Ann. Math.} {\bf 57} (1953) 591.
\bibitem{henteit} M. Henneaux and C. Teitelboim, {\em
Quantization of
Gauge Systems}, Princeton University Press (Princeton: 1992).
\end{thebibliography}
\end{document}